\PassOptionsToPackage{unicode}{hyperref}
\PassOptionsToPackage{hyphens}{url}
\PassOptionsToPackage{dvipsnames,svgnames,x11names}{xcolor}
\documentclass[12pt]{article}

\usepackage{amsmath,amssymb}
\usepackage{iftex}
\ifPDFTeX
  \usepackage[T1]{fontenc}
  \usepackage[utf8]{inputenc}
  \usepackage{textcomp} 
\else 
  \usepackage[T1]{fontenc}
  \usepackage[utf8]{inputenc}
  \usepackage{lmodern}
\fi
\IfFileExists{upquote.sty}{\usepackage{upquote}}{}
\IfFileExists{microtype.sty}{
  \usepackage[]{microtype}
  \UseMicrotypeSet[protrusion]{basicmath} 
}{}
\makeatletter
\@ifundefined{KOMAClassName}{
  \IfFileExists{parskip.sty}{%
    \usepackage{parskip}
  }{
    \setlength{\parindent}{0pt}
    \setlength{\parskip}{6pt plus 2pt minus 1pt}}
}{
  \KOMAoptions{parskip=half}}
\makeatother
\usepackage{xcolor}
\makeatletter
\ifx\paragraph\undefined\else
  \let\oldparagraph\paragraph
  \renewcommand{\paragraph}{
    \@ifstar
      \xxxParagraphStar
      \xxxParagraphNoStar
  }
  \newcommand{\xxxParagraphStar}[1]{\oldparagraph*{#1}\mbox{}}
  \newcommand{\xxxParagraphNoStar}[1]{\oldparagraph{#1}\mbox{}}
\fi
\ifx\subparagraph\undefined\else
  \let\oldsubparagraph\subparagraph
  \renewcommand{\subparagraph}{
    \@ifstar
      \xxxSubParagraphStar
      \xxxSubParagraphNoStar
  }
  \newcommand{\xxxSubParagraphStar}[1]{\oldsubparagraph*{#1}\mbox{}}
  \newcommand{\xxxSubParagraphNoStar}[1]{\oldsubparagraph{#1}\mbox{}}
\fi
\makeatother

\usepackage{longtable,booktabs,array}
\usepackage{calc} 
\usepackage{etoolbox}
\makeatletter
\patchcmd\longtable{\par}{\if@noskipsec\mbox{}\fi\par}{}{}
\makeatother
\IfFileExists{footnotehyper.sty}{\usepackage{footnotehyper}}{\usepackage{footnote}}
\makesavenoteenv{longtable}
\usepackage{graphicx}
\makeatletter
\def\maxwidth{\ifdim\Gin@nat@width>\linewidth\linewidth\else\Gin@nat@width\fi}
\def\maxheight{\ifdim\Gin@nat@height>\textheight\textheight\else\Gin@nat@height\fi}
\makeatother
\setkeys{Gin}{width=\maxwidth,height=\maxheight,keepaspectratio}
\makeatletter
\def\fps@figure{htbp}
\makeatother

\makeatletter
\@ifpackageloaded{caption}{}{\usepackage{caption}}
\AtBeginDocument{%
\ifdefined\contentsname
  \renewcommand*\contentsname{Table of contents}
\else
  \newcommand\contentsname{Table of contents}
\fi
\ifdefined\listfigurename
  \renewcommand*\listfigurename{List of Figures}
\else
  \newcommand\listfigurename{List of Figures}
\fi
\ifdefined\listtablename
  \renewcommand*\listtablename{List of Tables}
\else
  \newcommand\listtablename{List of Tables}
\fi
\ifdefined\figurename
  \renewcommand*\figurename{Figure}
\else
  \newcommand\figurename{Figure}
\fi
\ifdefined\tablename
  \renewcommand*\tablename{Table}
\else
  \newcommand\tablename{Table}
\fi
}
\@ifpackageloaded{float}{}{\usepackage{float}}
\floatstyle{ruled}
\@ifundefined{c@chapter}{\newfloat{codelisting}{h}{lop}}{\newfloat{codelisting}{h}{lop}[chapter]}
\floatname{codelisting}{Listing}

\makeatother
\makeatletter
\@ifpackageloaded{caption}{}{\usepackage{caption}}
\@ifpackageloaded{subcaption}{}{\usepackage{subcaption}}
\makeatother

\ifLuaTeX
  \usepackage{selnolig}  
\fi
\usepackage[authoryear]{natbib}
\usepackage{bookmark}

\IfFileExists{xurl.sty}{\usepackage{xurl}}{} 
\definecolor{CardinalRed}{cmyk}{0,1,0.65,0.34} 
\hypersetup{
  pdftitle={Title},
  pdfauthor={Author 1; Author 2},
  pdfkeywords={3 to 6 keywords, that do not appear in the title},
  colorlinks=true,
  linkcolor={CardinalRed},
  filecolor={CardinalRed},
  citecolor={CardinalRed},
  urlcolor={CardinalRed},
  pdfcreator={LaTeX via pandoc}}

\newcommand{\anon}{1}

\usepackage{amsthm}
\usepackage{amssymb}
\usepackage{amsmath}

\theoremstyle{definition}
\newtheorem{assumption}{Assumption}
\newtheorem*{theorem*}{Theorem}

\newtheorem*{rmk*}{remark}
\newtheorem{proposition}{Proposition}
\newtheorem{lemma}{Lemma}

\newtheorem{definition}{Definition}
\newtheorem{remark}{Remark}
\newtheorem{corollary}{Corollary}
\newtheorem*{corollary*}{Corollary}

\usepackage{bbm}
\usepackage{enumerate}
\usepackage{bm}

\usepackage{tikz}
\usetikzlibrary{positioning}
\newcommand{\mcirc}{\tikz \draw (0,0) circle[radius=1.1mm];}

\usepackage{placeins}
\usepackage{multirow}

\graphicspath{ {./Graphs/} }
\def\correct{correctly specified}
\def\xb{\bar X}
\def\didsc{{\textsc{did}}}

\def\gtgp{g \to g'}
\def\tdg{\tau_{\didsc, \gtgp}}
\def\htdgi{\htau_{\didsc, \gtgp,*}}
\def\htdgix{\htdgi(x)}

\def\htdga{\htau_{\didsc, \gtgp,+}}
\def\htdgax{\htdga(x)}

\def\tdgx{\tdg(x)}

\def\teg{\tau_{\textup{em},\gtgp}}
\def\tegx{\teg(x)}
\def\tcg{\tau_{\cm,\gtgp}}
\def\tcgx{\tcg(x)}

\def\ticg{\tau_{i,\cm, \gtgp}}

\def\tdxg{\tau_{\didx,\gtgp}}
\def\htdxgi{\htau_{\didx, \gtgp, *}}
\def\htdxga{\htau_{\didx, \gtgp, +}}
\def\texg{\tau_{\emx,\gtgp}}

\def\tcg{\tau_{\cm, \gtgp}}

\def\cptf{canonical parallel trends}
\def\cfptf{canonical and factorial parallel trends}

\def\edyijgx{\E[\dyij\mid \ggxxs ]}
\def\itp{{i_t}}
\def\ait{\alpha_{it}}
\def\itjt{{i_tjt}}
\def\itj{{i_tj}}

\def\ist{{i_t}}
\def\istj{{i_tj}}
\def\epij{\ep_{ij}}

\def\yitjt{Y_\ijt}
\def\xitj{X_\ijt}

\def\ggxx{\gig, \xix}
\def\ggxxs{\gig, \xijx}
\def\edygx{\E[\dyi \mid \ggxx]}
\def\eygx{\E[\yit \mid \ggxx]}

\def\ds{\displaystyle}
\def\hdy{\widehat{\dy}}
\def\hdya{\widehat{\dy_+}}
\def\hdyagx{\hdya(g,x)}
\def\hdyi{\widehat{\dy_*}}
\def\hdyigx{\hdyi(g,x)}

\def\tdxx{\tdid(X_i)}

\def\tdxf{\tdid(x)}

\def\prop{Proposition}
\def\ddef{Definition}

\def\xxi{\xx_i}
\def\yia{Y_\post}

\def\sec{Section}

\def\sl{\su\ level}
\def\slh{\su-level}
\def\sld{\slh\ data}
\def\sla{\slh\ analysis}

\def\sul{\sl}
\def\sulh{\slh}
\def\suld{\sld}
\def\sula{\sla}
\def\ulh{unit-level}

\def\ijt{{ijt}}
\def\yijt{Y_{\ijt}}
\def\xijt{X_{\ijt}}

\def\gij{G_{ij}}
 
\def\git{G_{it}}

\def\xij{X_{ij}}
\def\yijpost{Y_{ij, \post}}
\def\yijpre{Y_{ij, \pre}}
\def\dyij{\Delta Y_{ij}}

\def\su{subunit}

\def\rc{RCS}

\def\pps{\pp\ scheme}

\def\prf{panel-\rc}
\def\pp{panel-panel}
\def\ppp{Panel-panel}
\def\ppr{Panel-\rc}
\def\rr{\rc-\rc}

\def\dy{\Delta Y}

\newcommand{\mbr}{\mathbb R}

\newcommand{\ind}[1]{1_{\{#1\}}}

\def\lsf{OLS fit}

\def\olsa{OLS$_+$}
\def\olsi{OLS$_*$}

\def\olsas{OLS$^\prime_{+}$}
\def\olsis{OLS$^\prime_{*}$}

\def\twa{TWFE$_+$}
\def\twi{TWFE$_*$}

\def\twas{TWFE$^\prime_+$}
\def\twis{TWFE$^\prime_{*}$}
\def\twfea{\twa}
\def\twfei{\twi}

\def\mx{\mathcal X}
\def\ximx{x\in\mx}

\def\hboi{\hb_{1,*}}
\def\hbgi{\hb_{G,*}}
\def\hbxi{\hb_{X,*}}
\def\hbgxi{\hb_{GX,*}}
\def\hbxit{\hbxi^\T}
\def\hbgxit{\hbgxi^\T}

\def\hboa{\hb_{1,+}}
\def\hbxa{\hb_{X,+}}
\def\hbga{\hb_{G,+}}

\def\mg{\mathcal G}
\def\gimg{g\in\mg}
\def\gig{\gi = g}
\def\gio{\gi = 1}
\def\giz{\gi = 0}

\def\xix{X_i = x}
\def\xijx{\xij = x}

\def\gigp{\gi = g'}

\def\cmf{causal moderation}

\def\cm{\textup{cm}}

\def\hb{\hat\beta}

\def\zi{Z_i}
\def\gi{G_i}
\def\xxi{X_i}

\def\tds{\tdid'}
\def\tdsxf{\tdsx}
\def\tdsx{\tds(x)}

\def\tdx{\tdidx}
\def\tdidx{\tau_\didx}
\def\didx{{\textup{\textsc{did}-x}}}

\def\htdxa{\htau_{\textup{\textsc{did}-x},+}}
\def\htdax{\htda(x)}
\def\htda{\htau_{\textsc{did},+}}
\def\htdaxx{\htda(X_i)}

\def\htdxi{\htau_{\textup{\textsc{did}-x},*}}
\def\htdix{\htdi(x)}
\def\htdi{\htau_{\textsc{did},*}}
\def\htdixx{\htdi(X_i)}

\def\assmc{Assumptions~\ref{assm:ae}--\ref{assm:pta} and \ref{assm:fpt}}
\def\tg{\tau_{G \mid Z = 1}}
\def\E{\mathbb E}
\usepackage{pifont}

\def\ep{\epsilon}

\def\tc{\tau_\textup{cm}}

\def\tex{\tau_\emx}
\def\emx{{\textup{em-x}}}

\def\htdx{\htau_{\textup{\textsc{did}-x}}}
\def\ip{1_{\{t = \post\}}}

\def\assmerf{\assmer~(exclusion restriction)}
\def\assmer{Assumption~\ref{assm:er}}
\def\assmccfpt{Assumptions \ref{assm:cpt}--\ref{assm:cfpt} (conditional canonical and factorial parallel trends)}
\def\assmna{Assumption~\ref{assm:na} (no anticipation)}

\def\assmae{Assumption~\ref{assm:ae}}
\def\assmaef{Assumption~\ref{assm:ae}~(universal exposure)}

\def\assmol{Assumption~\ref{assm:overlap}}

\def\assmpt{Assumption~\ref{assm:pta}}
\def\assmptf{\assmpt~(canonical parallel trends)}

\def\assmcpt{Assumption~\ref{assm:cpt} (conditional canonical parallel trends)}
\def\assmcfpt{Assumption~\ref{assm:cfpt} (conditional factorial parallel trends)}

\def\ee{\mathbb E}

\def\tdid{\tau_\didsc}

\newcommand{\sumi}{\sum_{i=1}^n}
\newcommand{\meani}{n^{-1}\sum_{i=1}^n}
\def\T{\top}

\newcommand{\ot}[1]{1, \ldots, #1}

\newcommand{\indep}{{\perp\!\!\!\perp}}
\def\assmott{Assumptions \ref{assm:ae}--\ref{assm:pta}}

\def\plim{\textup{plim}\,}
\def\epi{\epsilon_i}
\def\yit{Y_{it}}
\def\htd{\htau_\textsc{did}}

\def\dyi{\Delta Y_i}

\def\tipw{\tau_\textsc{ipw}}
\def\tipwo{\tau_{\ipw\mid G=1}}
\def\tipwz{\tau_{\ipw\mid G=0}}

\def\ipw{{\textsc{ipw}}}

\def\tatt{\tau_{\textup{att}}}

\def\tic{\tau_{i,\textup{cm}}}

\def\tigz{\tau_{i, G \mid Z = z}}
\def\tizgg{\tau_{i, Z \mid G}}

\def\tizgo{\tau_{i, Z \mid G = 1}}
\def\tizgz{\tau_{i, Z \mid G = 0}}
\def\tizg{\tau_{i, Z \mid G = g}}
\def\tizgp{\tau_{i, Z\mid G = g'}}

\def\tgzo{\tau_{G \mid Z=1}}

\def\te{\tau_\textup{em}}
\def\tem{\te}

\def\tc{\tau_{\textup{cm}}}

\def\htau{\hat\tau}

\def\after{\textup{post}}
\def\before{\textup{pre}}

\def\pre{\textup{pre}}
\def\post{\textup{post}}

\def\yia{Y_{i, \after}}
\def\yib{Y_{i, \before}}

\def\pr{\mathbb{P}}

\def\begina{\begin{eqnarray*}}
\def\enda{\end{eqnarray*}}
\def\begine{\begin{enumerate}}
\def\ende{\end{enumerate}}

\def\begini{\begin{itemize}}
\def\endi{\end{itemize}}

\def\beginy{\begin{eqnarray}}
\def\endy{\end{eqnarray}}

\newcommand{\cls}[1]{\\\centerline{$#1$}} 

\def\begineq{\begin{equation}}
\def\endeq{\end{equation}}

\def\bo{\beta_1}
\def\bg{\beta_G}
\def\bx{\beta_X}
\def\bxt{\bx^\T}
\def\bgxt{\beta_{GX}^\T}
\def\bgx{\beta_{GX}}

\begin{document}
\def\spacingset#1{\renewcommand{\baselinestretch}%
{#1}\small\normalsize} \spacingset{1}

\def\ttl{Factorial Difference-in-Differences}

\if1\anon
{
\date{
\normalsize(Forthcoming, \emph{
Journal of the American Statistical Association})
}
\title{\bf \ttl%
  \thanks{
    Yiqing Xu and Anqi Zhao contribute equally to this paper. Yiqing Xu, Assistant Professor, Department of Political Science, Stanford University. Email: \url{yiqingxu@stanford.edu}. Anqi Zhao, Assistant Professor, Fuqua School of Business, Duke University. Email: \url{anqi.zhao@duke.edu}. Peng Ding, Associate Professor, Department of Statistics, University of California, Berkeley. Email: \url{pengdingpku@berkeley.edu}. 
    }}
  \author{Yiqing Xu\\
    (Stanford) \and 
    Anqi Zhao\\ (Duke) \and
    Peng Ding\\ (UC Berkeley)}
  \maketitle
} \fi

\if0\anon
{
  \bigskip
  \bigskip
  \bigskip
  \begin{center}
    {\LARGE\bf \ttl}
\end{center}
  \medskip
} \fi

\addtocounter{page}{-1}
\thispagestyle{empty}

\bigskip
\begin{abstract}\noindent 
We formulate \emph{factorial difference-in-differences} (FDID), a research design that extends canonical difference-in-differences (DID) to settings in which an event affects all units. In many panel data applications, researchers exploit cross-sectional variation in a baseline factor alongside temporal variation in the event, but the corresponding estimand is often implicit and the justification for applying the DID estimator remains unclear. We frame FDID as a factorial design with two factors, the baseline factor $G$ and the exposure level $Z$, and define effect modification and causal moderation as the associative and causal effects of $G$ on the effect of $Z$, respectively. Under standard DID assumptions of no anticipation and parallel trends, the DID estimator identifies effect modification but not causal moderation. Identifying the latter requires an additional \emph{factorial parallel trends} assumption, that is, mean independence between $G$ and potential outcome trends. We extend the framework to conditionally valid assumptions and regression-based implementations, and further to repeated cross-sectional data and continuous $G$. We demonstrate the framework with an empirical application on the role of social capital in famine relief in China.

\bigskip\noindent%
{\it Keywords:}  difference-in-differences, factorial design, panel data, parallel trends 
\end{abstract}

\vfill

\newpage
\spacingset{1.8} 

\section{Introduction}\label{sec:intro}
Social science research often relies on panel data to establish causality. One common approach involves exploiting cross-sectional variation in a baseline factor $G$ and temporal variation in exposure to a common event affecting all units, and applying the difference-in-differences (DID) estimator in a panel setting. As our running example, \citet{cao2022clans} examine how social capital ($G$), measured by the density of genealogy books, mitigated the mortality surge during China's Great Famine from 1958 to 1961 (the event), using a county-year panel. The authors interpret the coefficient of the interaction term between $G$ and an indicator of the famine years from a two-way fixed effects (TWFE) regression as the causal effect of social capital on famine relief, and describe their approach as a DID method. Section~\ref{sec:examples} gives details on this study and five additional examples that employ a similar approach.

Although the coefficient from a TWFE regression is numerically equal to a DID estimate and this approach is often referred to as a DID method in the empirical literature, it differs from the canonical DID popularized by \citet{card} because it lacks a clean control group unexposed to the event. The corresponding causal interpretation of the DID estimator in such settings is absent from the methodological literature, and empirical applications using this approach lack clear definitions of target causal estimands.

This paper aims to bring conceptual clarity to this empirical approach, which we term {\it factorial difference-in-differences} (FDID). We define FDID as a research design, or an identification strategy, that employs the DID estimator to recover interpretable quantities of interest using observations before and after a one-time event that affects all units, provided clearly stated identifying assumptions hold. Here, we highlight the distinction between the DID estimator and the canonical DID and FDID research designs. Applied to panel data, the DID estimator, denoted by $\htd$, calculates the difference in before-after differences with respect to an event between two groups, denoted by $G_i = 1$ and $G_i = 0$. In contrast, a research design encompasses not only the estimator but also the identifying assumptions and identification results. To simplify the presentation, we will refer to the FDID and canonical DID research designs as ``FDID'' and ``canonical DID,'' respectively, when no confusion is likely to arise. 

We present our main theoretical results for FDID in the two-group, two-period case. The key innovation is to augment the potential outcomes framework to include both the baseline factor $G$ and the exposure indicator $Z$, motivating the term ``factorial'' in FDID. This formulation clarifies what the probability limit of $\htd$, denoted by $\tdid$, identifies under different identification assumptions, and how these results relate to canonical DID.

In canonical DID, $\tdid$ identifies the average treatment effect on the treated (ATT) under the no anticipation and parallel trends assumptions \citep{angrist2009mostly}. In FDID, the focus shifts to two other estimands: effect modification and causal moderation \citep{tyler, bansak2020estimating}. Effect modification captures how the effect of $Z$ varies across groups defined by $G$, but does not represent $G$’s causal effect. In our running example, this corresponds to the statement that the mortality increase caused by the famine is smaller in counties with higher levels of social capital.  Causal moderation, by contrast, has a direct causal interpretation as $G$’s effect on the impact of $Z$, or symmetrically, $Z$’s effect on the impact of $G$. In our example, causal moderation corresponds to two equivalent statements: (i) social capital reduced the famine’s negative impact on mortality, as emphasized in the original paper, or (ii) social capital had a stronger effect on mortality during the famine than it would have otherwise. 
Our key result is that under the no anticipation and canonical parallel trends assumptions, $\tdid$ identifies effect modification, whereas recovering causal moderation requires additional assumptions. One such condition is the \emph{factorial parallel trends} assumption, that is, mean independence between $G$ and potential outcomes trends.  Intuitively, it holds in the absence of any unobserved confounder correlated with both $G$ and the outcome trends. For example, to identify the causal moderation of social capital on the famine’s effect on mortality, one must rule out any unobserved factor—such as income or governance quality—that is correlated with both social capital and changes in mortality between famine and non-famine years.

Moreover, we show that canonical DID can be reframed as a special case of FDID under an additional exclusion restriction requiring that exposure to the event has no effect on the outcome for units with $\giz$. Under this assumption, $G$'s effect modification simplifies to the average effect of $Z$ on units with $\gio$, analogous to the ATT in canonical DID. In our running example, this assumption implies that the famine had no impact on localities with low (or high) levels of social capital, an implausible claim in this setting.
Alternatively, if researchers assume that $G$ has no causal impact on the outcome in the absence of the event, then causal moderation reduces to $G$'s  average conditional effect given exposure to the event. With our example, this assumption implies that social capital would have no effect on mortality had the famine not occurred, which is possible but not suggested by the original paper.

We extend the framework to settings where canonical and factorial parallel trends hold only conditional on additional time-invariant covariates. We formalize identification results for the conditional DID estimator and clarify the assumptions needed to justify regression-based analysis. We also provide theoretical results for applications with repeated cross-sectional data and with a continuous baseline factor $G$.

Our contributions are twofold. First, we establish the causal interpretation of a widely used empirical approach in the social sciences, clarifying the identifying assumptions required to recover causal estimands of interest. This framework advances the discussion of causal panel analysis with the DID estimator and TWFE models---for recent reviews, see \citet{roth2023s}, \citet{chiu2023}, and \citet{imbens2023panel}. Second, we contribute to the literature on factorial designs \citep{tyler, bansak2020estimating, han2021contrast, pashley2023causal, yu2023balancing} by, to our knowledge, being the first to extend factorial designs to observational panel settings and to analyze the role of parallel trends assumptions in this context.

The rest of the paper is organized as follows. Section~\ref{sec:examples} gives six FDID examples appearing in the empirical literature. Section~\ref{sec:setup} formalizes FDID under the two-group, two-period panel case. Section~\ref{sec:identification} states identification results for FDID and reconciles FDID with canonical DID. Sections~\ref{sec:ext_cond}--\ref{sec:ext_rcs_G} discuss extensions to conditionally valid assumptions, repeated cross-sectional data, and continuous $G$. Section~\ref{sec:app} illustrates our theory with the running example. Section~\ref{sec:conclude} concludes. The Supplementary Materials provide technical details, and the replication files are available on GitHub at \url{https://github.com/xuyiqing/fdid_paper}.


\section{FDID Examples}\label{sec:examples}

In this section, we present six empirical examples from economics, political science, and finance that align with the FDID research design. In each case, researchers obtain key estimates using a TWFE regression and describe the approach as a DID method. However, their intended estimands often differ.

\citet{Squicciarini2020} examines whether Catholicism, proxied by the share of refractory clergy in 1791, hindered economic growth during the Second Industrial Revolution in France. One main analysis relies on a longitudinal dataset of French departments from 1866 to 1911. The baseline factor $G$ is the 1791 share of refractory clergy, and the event is the Second Industrial Revolution in the late 19th century (post-1870), to which all departments were presumably exposed. It is not entirely clear about which estimand the study aims to identify. The author argues that the results pointed to a causal interpretation of the relationship between religiosity and economic development \emph{during} the Second Industrial Revolution, which corresponds most closely to $G$’s average conditional effect. The study also describes itself as examining ``the differential diffusion of technical education and industrial development'' (p. 3455), which can be interpreted as targeting the effect modification of $G$. 

\citet{fouka2019how} studies ``the effect of taste-based discrimination on the assimilation decisions of immigrant minorities'' in the United States (abstract). The repeated cross-sectional data include all men born in the US from 1880 to 1930 to a German-born father, organized by state and birth year. The baseline factor $G$ is state-level measures of anti-Germanism, such as support for Woodrow Wilson in the 1916 presidential Election. The event is World War~I, which started in 1917. The outcome is a foreign name index. Although not stated explicitly, the intended estimand is likely either the causal moderation or $G$'s average conditional effect post World War I.  %

\citet{charnysh2022explaining} argues that during crises, states allocate fewer resources to less ``legible'' groups---those from which they cannot gather reliable information or effectively collect taxes. Using district-level panel data with yearly observations from Imperial Russia, the author studies the 1891–1892 Russian famine. The baseline factor $G$ is the district-level share of Muslims, and the event is the famine. The findings show that districts with larger Muslim populations experienced higher mortality rates during the famine. The author avoids explicit causal claims, so the analysis can be interpreted as targeting the effect modification of $G$.

\citet{chen2024powerholders} argue during a major state-building episode in ancient China, the number of aristocrats from prefectures recruited into the imperial bureaucracy increased more in localities with strong military presence. The study uses prefecture-level panel data from the Northern Wei Dynasty (384–534 CE). The baseline factor $G$ is whether a prefecture had fourth-century military strongholds, and the event is a state-building reform initiated by Empress Dowager Feng (477–490 CE). The authors describe their empirical strategy as a ``canonical DD strategy,'' which ``relies on the parallel-trend assumption to adopt a causal interpretation.'' Given the use of explicit causal language, the intended estimand is likely either the causal moderation or the average conditional effect of $G$.

\citet{chen2023pledgeability} examine how the ability to use corporate bonds as collateral (pledgeability) affects their prices, leveraging a policy change in Chinese bond markets. The study uses panel data on daily bond prices. The baseline factor $G$ is bond ratings, and the event is a policy change on December 8, 2014, when Chinese policymakers prohibited bonds rated below AAA from being used as collateral. The intended estimand is the ATT, the policy effect on AA and AA+ bonds. This setting reduces to canonical DID because the exclusion restriction---i.e., no policy effect on AAA and AA$-$ bonds---is plausible, given that AA$-$ bonds were already ineligible for pledging before the policy change.

\begin{figure}[!ht]
    \begin{center}
    \includegraphics[width=0.7\linewidth]{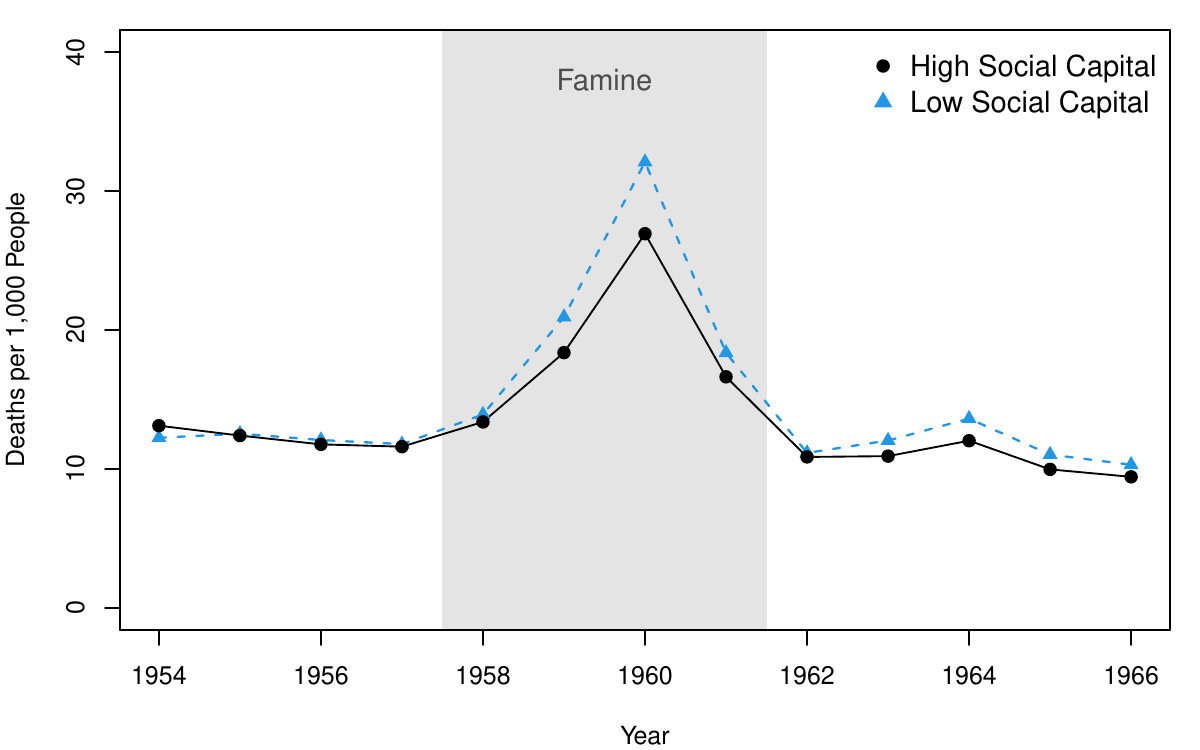}
    \end{center}\spacingset{1.2}    
    \caption{Average mortality rate across counties with high and low social capital. This figure resembles Figure 5(a) in \citet{cao2022clans}, although we use a balanced panel, which represents a subset of the data used by the authors.}
    \label{fig:raw}   
\end{figure}

Finally, \citet{cao2022clans} is our running example. Recall from \sec~\ref{sec:intro} that the baseline factor $G$ is social capital, and the event is China's Great Famine from 1958 to 1961. The authors intend to estimate the causal moderation of social capital on the famine's impact on mortality. In Section~\ref{sec:app}, we reanalyze this application using a balanced panel of 921 counties spanning the years 1954 to 1966. 
Figure~\ref{fig:raw} displays the average mortality rates during this period for two types of counties in the sample: those with high social capital and those with low social capital. While the average mortality rate rose sharply during the famine years in both groups of counties, the increase was noticeably higher in counties with low social capital compared with those with high social capital. 

\section{FDID in Two-Group, Two-Period Panel Setting}\label{sec:setup}

We present our main theoretical results in the two-group, two-period panel setting. This section introduces the notation, data structure, and DID estimator, and then defines the potential outcomes that form the basis for the estimands.

\subsection{Observed data and FDID setting} 

Assume a standard two-group, two-period panel setting with a one-time event and a study sample of $n$ units, indexed by $i = 1, \ldots, n$. For each unit $i$, we observe a binary baseline factor $\gi \in \{0,1\}$, an outcome measured at two time points—before and after the event—denoted by $\yib \in \mbr$ and $\yia \in \mbr$, and an exposure indicator $\zi \in \{0,1\}$, where $\zi = 1$ if unit $i$ is exposed when the event occurs and $\zi = 0$ otherwise. A defining feature of FDID is that all units are exposed to the event, so $Z_i = 1$ for all $i$, as formalized in Assumption~\ref{assm:ae} and Definition~\ref{def:FDID_setting} below.

\begin{assumption}[Universal exposure]\label{assm:ae} $\zi = 1$ for all $i =\ot{n}$.
\end{assumption}

\begin{definition}[FDID setting]\label{def:FDID_setting}
The two-group, two-period (2$\times$2) panel setting for FDID consists of the 2$\times$2 panel data $\{(\gi, \yib, \yia, \zi): G_i \in \{0,1\}\}_{i=1}^n$ and Assumption \ref{assm:ae}.
\end{definition}

In the FDID setting, the exposure indicator $\zi$ equals one for all units and thus appears redundant. However, it is essential for defining the potential values of $\yib$ and $\yia$ that would have been observed in the absence of the event. These potential outcomes provide the basis for defining the causal estimands and stating the identification assumptions in FDID. A similar use of $\zi$ appears in \cite{holland1986research} to clarify Lord's paradox in settings identical to FDID. The departure from canonical DID arises from the absence of a one-to-one mapping between $G$ and $Z$.

\subsection{DID estimator}

Let $\dyi = \yia - \yib$ denote the before-after difference in the outcomes of unit $i$. The DID estimator is the difference in the average $\dyi$ between the two groups defined by the baseline factor $G$: 
\begin{equation}\label{eq:htd}
\htd  =  n_1^{-1}\sum_{i: \gio} \dyi - n_0^{-1}\sum_{i: \giz} \dyi, 
\end{equation} 
where $n_g$ is the number of units with $\gig$. We assume throughout that units are drawn from a common population distribution. Define
\begin{equation}\label{eq:tdid}
\tdid = \E[\dyi \mid \gio ] - \E[\dyi \mid \giz  ] 
\end{equation}
as the probability limit of $\htd$ as $n \to \infty$, commonly referred to as the DID estimand. Our goal is to clarify the causal interpretation of $\tdid$ in the FDID setting under various identifying assumptions.

\begin{remark}\label{rmk:cross-sec}
From \eqref{eq:htd}--\eqref{eq:tdid}, the DID estimator and estimand, $(\htd, \tdid)$, depend on the panel outcomes $(\yib, \yia)$ only through their difference $\dyi$, and equal the difference-in-means estimator and estimand based on the cross-sectional data ${(G_i, \dyi)}_{i=1}^n$. This equivalence between DID and cross-sectional analyses underlies the key identifying assumptions that justify the causal interpretation of $\tdid$, as discussed in \sec~\ref{sec:identification}.
\end{remark}

\subsection{Potential outcomes}\label{sec:po}

\begin{figure}[!ht]
\vspace{-1em}
 \begin{center}
 \includegraphics[width = 0.6\textwidth]{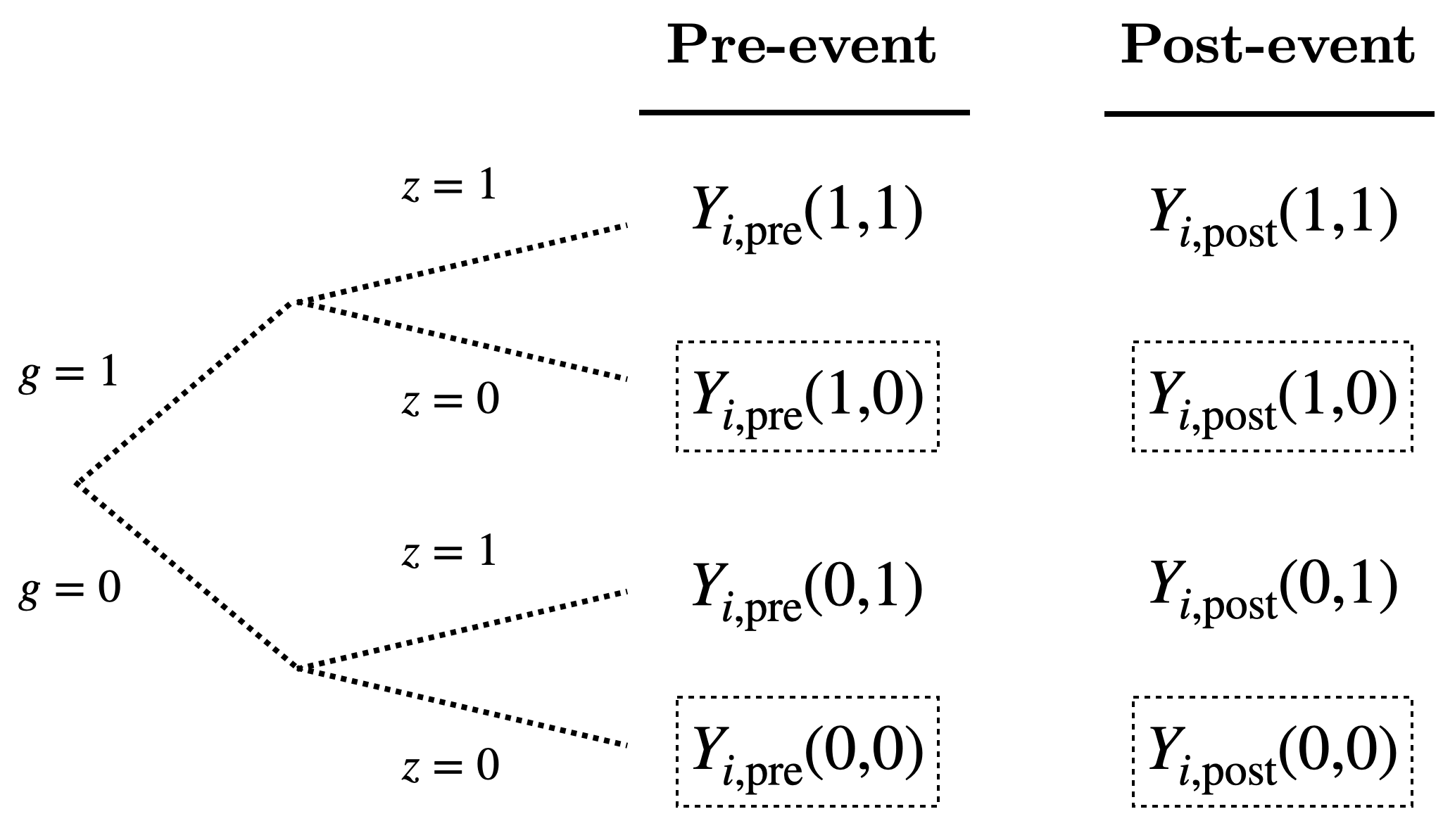}
 \end{center} 
\spacingset{1.2}\caption{\label{fig:po} Potential outcomes under FDID. The four potential outcomes with $z = 0$ in dotted boxes are unobservable for any unit in the FDID setting.}\medskip
\end{figure}

We now define the potential outcomes under FDID. Unlike the classic DID framework, we define them with respect to both the baseline factor $\gi$ and the exposure level $\zi$. For $t \in \{\pre,\post\}$, let $\yit(g,z)$ denote the potential value of $\yit$ if $G$ were set at $g$ and $Z$ were set at $z$ for unit $i$. Let $\ind{\cdot}$ be the indicator function. The observed outcome satisfies $\yit = \sum_{g, z=0,1} \ind{(\gi,\zi) = (g,z)} \yit(g,z)= \yit(\gi, \zi)$, which reduces to $\yit =  \yit(\gi, 1)$ under Assumption \ref{assm:ae}. Thus, the four potential outcomes with $z=0$, $\{\yit(g,0): t \in \{\pre,\post\},\ g \in \{0,1\}\}$, are unobservable for all units in the FDID setting. Figure~\ref{fig:po} illustrates this setup.

\subsection{Estimands}

We now formalize four estimands that $\tdid$ can identify in the FDID setting under different assumptions. Following the literature on causal inference with factorial experiments \citep{td, bansak2020estimating, zdfact}, we begin by defining three unit-level effects. Let $\tizg = \yia(g, 1) - \yia(g, 0)$ denote the effect of exposure on unit $i$ if the baseline factor $G$ were at level $g$. Let $\tigz = \yia(1, z) - \yia(0, z)$
denote the effect of the baseline factor $G$ on unit $i$ if its exposure level were at $z$. 
Define
\cls{
\tic = \tizgo  - \tizgz 
}
as the {\it \cmf} of $G$ on the effect of exposure for unit $i$, capturing how the effect of exposure differs between the two levels of $G$.
Note that 
\cls{
\tic = \yia(1,1) - \yia(1,0) - \yia(0,1) + \yia(0,0) = \tau_{i, G\mid Z = 1} - \tau_{i, G\mid Z=0}. 
}
Thus, $\tic$ is symmetric in $G$ and $Z$, and can also be interpreted as the causal moderation of $Z$ on the effect of $G$ for unit $i$. In the literature \citep[see, e.g.,][]{tyler}, $\tic$ is also referred to as the {\it interaction} between $G$ and $Z$. 

\ddef~\ref{def:estimands1} below formalizes effect modification and causal moderation as comparisons of the exposure effect, $\tau_{i, Z\mid G=g}$, between the two levels of $G$, extending \cite{tyler} and \cite{bansak2020estimating} to the panel setting. To highlight the distinction between these concepts, define $\tizgg $ as the value of $\tau_{i, Z| G =g}$ when the baseline factor $\gi$ is at its observed level, i.e., 
\cls{
\tizgg = \left\{
\begin{array}{ll}
\tizgo & \text{if $\gio $};\\
\tizgz & \text{if $\giz  $}.
\end{array}
\right.
}

\begin{definition}[Effect modification and causal moderation]\label{def:estimands1}
Define\\
\centerline{$\te = \E[\tizgg  \mid \gio ] - \E[\tizgg  \mid \giz  ] = \E[\tizgo  \mid \gio ] - \E[\tizgz  \mid \giz  ]$}
as the {\it effect modification} of the baseline factor $G$ on the effect of exposure;\\
\centerline{$\tc= \E[\tic] = \E[\tizgo -\tizgz]$}
\noindent 
as the {\it causal moderation} of the baseline factor $G$ on the effect of exposure.
\end{definition}

Both $\te$ and $\tc$ capture heterogeneity in the effect of exposure, $\tizg$, across the two levels of the baseline factor $G$, but they emphasize different comparisons. On the one hand, $\tc$ compares $\tau_{i, Z\mid G=1}$ and $\tau_{i, Z\mid G=0}$ across all units and is conventionally regarded as a causal quantity \citep{holland1986research, frangakis2002principal, CausalImbens}. It addresses the causal question: Would the effect of exposure change if one intervened on $G$?  In our running example, this corresponds to whether the impact of exposure to China's Great Famine would have differed if a locality had randomly developed social capital, perhaps due to the migration of a large kinship clan. By contrast, $\te$ compares the average of $\tau_{i, Z\mid G=1}$ among units with $\gio$ to the average of $\tau_{i, Z\mid G=0}$ among units with $\giz$, each evaluated at the observed value of $\gi$.  Accordingly, $\te$ describes differences in exposure effects between the two groups of units defined by $G$, but does not address the same causal question as $\tc$. See \cite{tyler} for further discussion in the cross-sectional setting.

\begin{remark}\label{rmk:att}
In some applications, $G$ is an inherent characteristic of the unit, such as geography of a locality or race of a person, which cannot be meaningfully manipulated. In these cases, the augmented potential outcomes $Y_{i,t}(g, z)$ are not well defined at the counterfactual level of $G$, and thus $\tc$ is not a coherent causal estimand. Instead, $\te$ is a more relevant target. In addition, the causal moderation $\tc$ is the average of $\tic$ across all units, analogous to the average treatment effect. One may also consider $\E[\tic \mid \gig]$, the average causal moderation among units with $\gi = g$, analogous to the ATT. 
\end{remark}

\ddef~\ref{def:estimands2} below formalizes two conditional causal effects of $G$. 

\begin{definition}[Conditional causal effects of $G$]\label{def:estimands2}
Define 
\cls{\tatt = \E[\tizgg  \mid \gio ] = \E[\tizgo  \mid \gio ]}
as the average causal effect of exposure on units with $\gio$, and 
\cls{\tg = \E[\tau_{i, G\mid Z = 1}]} as the average causal effect of the baseline factor $G$ conditional on exposure.
\end{definition}

From \ddef~\ref{def:estimands2}, $\tatt$ is analogous to the ATT in canonical DID, with units satisfying $\gio$ viewed as the treated group. The subscript ``att'' reflects this connection. The estimand $\tg$ is the average causal effect of $G$ conditional on exposure to the event. Figure~\ref{fig:estimands} summarizes the relationships among these estimands, serving as a roadmap for our identification results.

\begin{figure}[!th]
    \centering
    \resizebox{0.7\textwidth}{!}{
    \begin{tikzpicture}
  \draw[thick] (-.5,1) rectangle (10.5, 6);

  \node at (.5, 4) {$\tdid$};
  \node at (5, 4) {$\tem$};
  \node at (9.5, 4) {$\tc$};
  \node at (5, 1.5) {$\tatt$}; 
  \node at (9.5, 1.5) {$\tau_{G\mid Z=1}$}; 

  \node[align=center] at (2.75, 5) {
  \renewcommand{\arraystretch}{.6}\footnotesize
  \begin{tabular}{c}Universal exposure\\No anticipation\\Parallel trends\end{tabular}};
  \node[align=center] at (7.25, 5) {
  \renewcommand{\arraystretch}{.6}\footnotesize
  \begin{tabular}{c}No anticipation\\Parallel trends\\Factorial parallel trends\end{tabular}};
  \node[align=center] at (3, 2.5) {
  \renewcommand{\arraystretch}{.6}\footnotesize
  \begin{tabular}{c}
  Exclusion restriction \\on $Z$ in $\{i : \giz  \}$
  \end{tabular}};
  \node[align=center] at (7.5, 2.5) {
  \renewcommand{\arraystretch}{.6}\footnotesize
  \begin{tabular}{c}
  Exclusion restriction \\on $G$ absent the event
  \end{tabular}};
  \draw[thick] (1.5, 4.05) -- (4, 4.05);
  \draw[thick] (1.5, 3.95) -- (4, 3.95);
  \draw[thick] (6, 4.05) -- (8.5, 4.05);
  \draw[thick] (6, 3.95) -- (8.5, 3.95);
  \draw[thick] (4.95, 3.5) -- (4.95, 2);
  \draw[thick] (5.05, 3.5) -- (5.05, 2);
  \draw[thick] (9.45, 3.5) -- (9.45, 2);
  \draw[thick] (9.55, 3.5) -- (9.55, 2);
\end{tikzpicture}
}\vspace{-1em}
\caption{Roadmap for key identification results \label{fig:estimands}}
\end{figure}

\FloatBarrier

\section{Identification}\label{sec:identification}

We now present the identification results. To preview, $\tdid$ identifies $\te$ under the canonical DID assumptions; identifying $\tc$, $\tatt$, and $\tg$ requires additional assumptions.

\subsection{Identification under canonical DID assumptions}\label{sec:tauem}

The observed pre-event outcome $\yib$ and its potential values $\{\yib(g,z): g,z  = 0, 1\}$ all occur before the event. A common, often implicit, assumption in the DID literature is that future events do not influence past potential outcomes, known as the \emph{no anticipation} assumption. We state this assumption explicitly in Assumption \ref{assm:na} below. It may be violated if units anticipate the event and adjust their behavior in advance.

\begin{assumption}[No anticipation]\label{assm:na}
$\yib(g, 0) = \yib(g, 1)$ for $i = \ot{n}$ and $g = 0,1$. 
\end{assumption}

Recall that $\dyi = \yia - \yib$ is the before-after difference in outcome for unit~$i$. Let $\dyi (g,z) = \yia(g,z) - \yib(g,z)$ denote the potential value of $\dyi$ if $G$ were set at $g$ and $Z$ were set at $z$ for unit $i$. Define $\dyi(\gi,0) = \yia(\gi,0)-\yib(\gi,0)$ as the potential before-after change for unit~$i$ under its observed baseline factor $G$ but assuming no exposure. Assumption \ref{assm:pta} below restates the canonical parallel trends assumption using the augmented potential outcomes.

\begin{assumption}[Canonical parallel trends]\label{assm:pta}
$\E[\dyi(\gi, 0) \mid \gio] = \E[\dyi( \gi, 0) \mid \gi =~0]$.
\end{assumption}
 
\begin{figure}[!ht]
 \vspace{-1em}\begin{center} 
 \includegraphics[width = 0.6\textwidth]{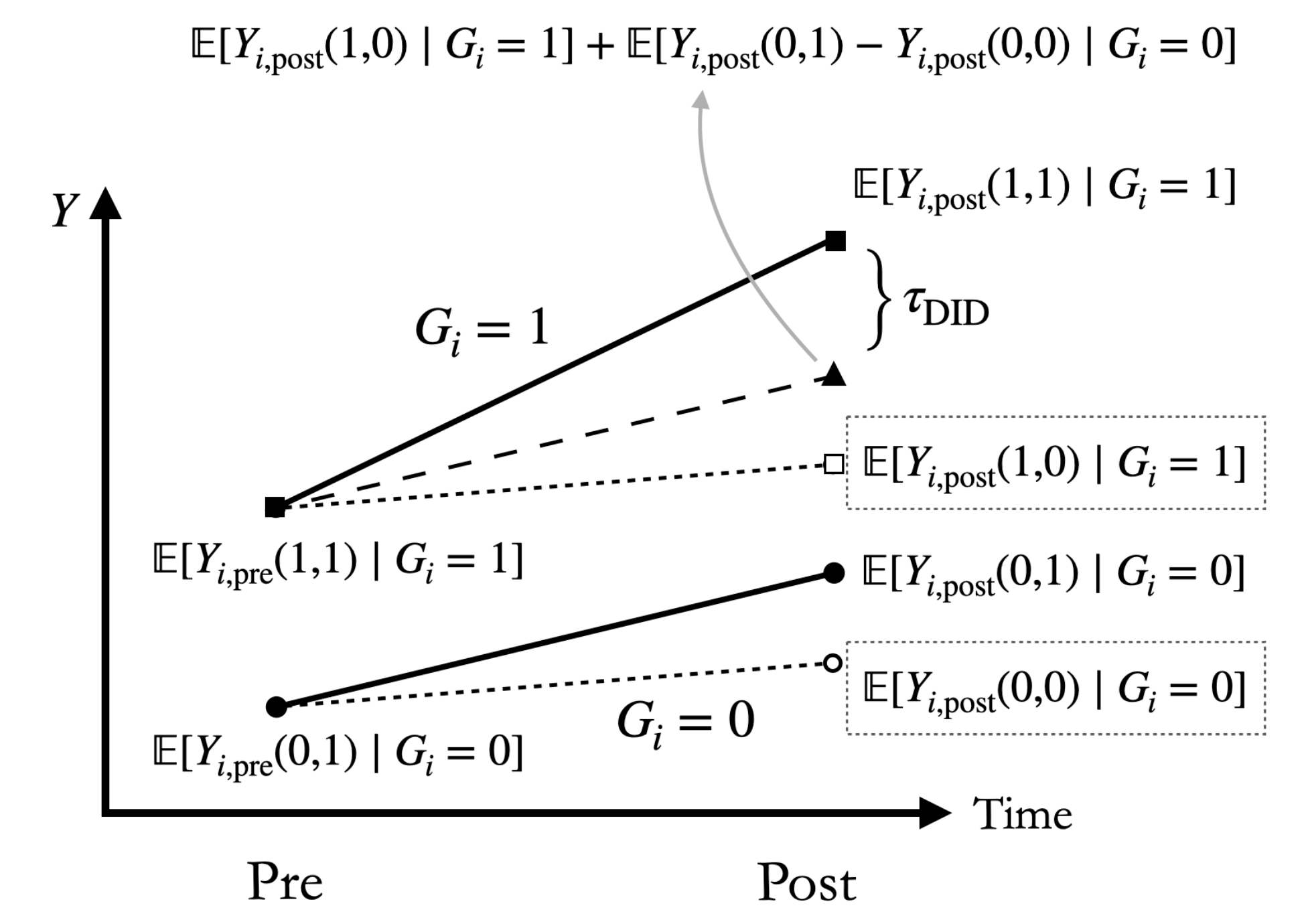} 
 \end{center} \vspace{-1em}
\spacingset{1.2}
\caption{\label{fig:id1} Identification result under FDID in Proposition~\ref{prop:tau_em}} \medskip
{\footnotesize
\textbf{Note:} Squares ($\blacksquare$ and $\square$) represent the group with $G_i = 1$, and circles ($\text{\ding{108}}$ and \mcirc) represent the group with $G_i = 0$. By {\assmna}, in the pre-event period, $\blacksquare = \E[\yib(1,0) \mid G_i = 1] = \E[\yib \mid G_{i} = 1]$ and $\text{\ding{108}} = \E[\yib(0,0) \mid G_i = 0] = \E[\yib \mid G_{i} = 0]$, so $\blacksquare$ and \ding{108} represent the pre- and post-event observed outcomes of the two groups. The long dashed line is drawn parallel to the solid line for $G_i = 0$ to represent the definition of the DID estimand, so in the post-period the distance between $\blacksquare$ and \ding{115} equals $\tdid$. This gives the first expression of $\text{\ding{115}}$: 
$$\text{\ding{115}} = \blacksquare - \tdid = \E[\yia(1,1)\mid G_i = 1] - \tdid.
$$ 
Under \assmpt, the two short dashed lines are also parallel. These parallel lines ensure that in the post period, the distance between \ding{115} and $\square$ equals the distance between \ding{108} and \mcirc.
This gives the second expression of $\text{\ding{115}}$:
$$
\text{\ding{115}} = \square + (\text{\ding{108}} - \text{\mcirc}) = \E[\yia(1,0)\mid G_i = 1] + \E[\yia(0,1) - \yia(0,0)\mid G_i = 0].
$$
Equating the two expressions for \ding{115} yields 
$$
\tdid = \E[\yia(1,1)-\yia(1,0)\mid G_i = 1] - \E[\yia(0,1) - \yia(0,0)\mid G_i = 0] = \te .
$$
}
\end{figure}

Assumption \ref{assm:pta} states that, in the absence of the event, the average change in outcome over time, $\dyi(\gi, 0)$, would be the same across the two groups defined by the baseline factor $G$. Together, Assumptions~\ref{assm:na}--\ref{assm:pta} form the canonical identifying assumptions in the DID literature, under which $\tdid$ identifies the ATT in the canonical DID setting. Proposition~\ref{prop:tau_em} below extends this classic result to the FDID setting and shows that, under these assumptions, $\tdid$ identifies $\te$. Figure~\ref{fig:id1} illustrates this identification result. 

\begin{proposition}\label{prop:tau_em}
If \assmott\ hold, then
$\tdid = \te$. 
\end{proposition}

\ddef~\ref{def:FDID_setting} and \prop~\ref{prop:tau_em} underscore two key differences between FDID and canonical DID. First, FDID assumes that all units are exposed to the event, whereas canonical DID relies on a clean, unexposed control group. Second, under the no anticipation and canonical parallel trends assumptions, the DID estimator identifies the ATT, a causal quantity, in canonical DID, but identifies $\te$, a descriptive quantity, in FDID. 
Despite these differences, we show below that the canonical DID research design can be reframed as a special case of FDID under an additional exclusion restriction assumption that the event has no effect on a group of units defined by $G$.

\begin{assumption}[Exclusion restriction]\label{assm:er}
$\E[\yia(0, 1) \mid \giz  ] = \E[\yia(0, 0) \mid \giz  ]$. 
\end{assumption}

{\assmer} ensures that the average post-event outcome of units with $\giz$ is unaffected by exposure. Conceptually, these units are exposed but unaffected, thus resembling the clean control group in canonical DID. Together, Assumption~\ref{assm:ae} (universal exposure) and \assmerf\ reproduce a canonical DID setting under FDID, with units satisfying $\gio $ and $\giz$ serving as the treated and control groups, respectively. This justifies interpreting $\tatt$ as the ATT analog in FDID. Moreover, {\assmer} implies $\E[\tizgz  \mid \giz  ] = 0$, so that $\te = \E[\tizgo \mid \gio ] - \E[\tizgz  \mid \giz  ] =\tatt$. This provides a causal interpretation of $\te$, as formalized in \prop~\ref{prop:recon} below.

\begin{proposition}\label{prop:recon}
(i) If \assmer\ holds, then $\te = \tatt$. (ii) If Assumptions \ref{assm:ae}--\ref{assm:er} hold, then $\tdid = \tatt$.
\end{proposition}

Recall that $\tatt$ is the ATT analog under FDID. \prop~\ref{prop:recon} reframes the classic identification result for the ATT in canonical DID within the FDID framework. \ddef~\ref{def:did} builds on this result and characterizes canonical DID as a special case of FDID. In Figure~\ref{fig:id1}, this corresponds to the post-period gaps between the solid triangle and hollow square, and between the solid circle and hollow circle, being closed.

\begin{definition}[Reframed canonical DID research design]\label{def:did}
The canonical DID setting is equivalent to the FDID setting in \ddef~\ref{def:FDID_setting} combined with \assmer, if the groups defined by $G$, $\{i:\gio\}$ and $\{i:\giz\}$, are viewed as the treatment and control groups, respectively. Under this setting, effect modification $\te$ reduces to the ATT analog $\tatt$, and is identified by $\tdid$ under Assumptions \ref{assm:na}--\ref{assm:pta}.
\end{definition}


\subsection{Identifying causal moderation and $G$'s conditional effect}\label{sec:tauc}

Propositions \ref{prop:tau_em}--\ref{prop:recon} establish the identification of $\te$ and $\tatt$ under FDID. In many applied studies, however, the primary quantity of interest is the causal moderation $\tc$ or the causal effect of $G$ given exposure, $\tgzo$; c.f.~\sec~\ref{sec:examples}. We discuss their identification below.

Recall that $\dyi (g,z) = \yia(g,z) - \yib(g,z)$ denotes the potential before-after change in outcome for unit~$i$. Assumption \ref{assm:fpt} below introduces a {\it factorial parallel trends} assumption, which requires mean independence between $\gi$ and $\dyi(g,z)$. 

\begin{assumption}[Factorial parallel trends]\label{assm:fpt}
$\E[\dyi(g, z) \mid \gio ] = \E[\dyi(g, z) \mid \giz  ]$ for $g,z = 0,1$. 
\end{assumption}

We call Assumption~\ref{assm:fpt} the factorial parallel trends assumption because, like canonical parallel trends (\assmpt), it requires equal average changes in potential outcomes across groups. 
However, now $\dyi(g,z)$ varies both $G$ and $Z$, implying mean independence between $G$ and all four potential outcome changes, $\dyi(g,z)$. By contrast, \assmpt\ requires only mean independence between $\gi$ and $\dyi(\gi,0)$, holding $G$ fixed at its observed value. 
A sufficient condition for Assumption~\ref{assm:fpt} is 
\begin{equation}\label{eq:indep}
G_i \indep \{\dyi(g,z):g, z = 0,1\}, 
\end{equation}
which states that $G$ is independent of the before-after changes in all four potential outcomes. As noted in Remark~\ref{rmk:cross-sec}, DID analysis with $(G_i,\yib,\yia)$ in FDID corresponds to cross-sectional analysis with $(G_i,\dyi)$. Analogously, condition~\eqref{eq:indep} resembles the standard random assignment assumption for $(G_i,\dyi,Z_i)$ in cross-sectional settings. 
Condition~\eqref{eq:indep} reflects the belief that differencing the potential outcomes removes confounding with respect to $G$. As noted earlier, some applied researchers do not recognize that Assumption~\ref{assm:fpt} or condition~(\ref{eq:indep}) is required to interpret DID estimates as causal moderation. Others, while not explicitly stating it, appear to have an intuitive understanding of this assumption and view it as more plausible than full random assignment $G_i \indep Y_i(g,z)$.%
\footnote{For example, \citet{fouka2019how} writes: ``I control for the potential time-varying effect of the share of the German population in the state, which is plausibly correlated with both (lower) support for Wilson and assimilation,'' and ``state-level anti-Germanism is potentially endogenous to pre-existing trends in German assimilation'' (p. 419), clearly acknowledging potential correlation between $G_i$ and $\dyi(g,z)$.} 
Nevertheless, the assumption, like unconfoundedness, is strong and untestable.

\prop~\ref{prop:tauc} below formalizes the conditions under which $\tc$ and $\te$ to coincide, thereby ensuring that $\tdid$ identifies $\tc$.

\begin{proposition}\label{prop:tauc}
(i) Under Assumptions \ref{assm:na}--\ref{assm:pta} and \ref{assm:fpt},  $\te = \tc$.  
(ii) Under \assmott\ and \ref{assm:fpt}, $\tdid = \te = \tc$.
\end{proposition}

\begin{remark}
The condition required for $\te = \tc$ is not unique. For example, mean independence between $\gi$ and $\{\tau_{i, Z\mid G=g}: g = 0,1\}$ also implies $\tc=\te$. We focus on Assumption \ref{assm:fpt} because  it is the one most often invoked, implicitly or explicitly, in empirical applications of FDID as researchers frequently argue that $G_{i}$ and $\dyi(g,z)$ are mean independent after conditioning on additional baseline covariates \citep[e.g.,][]{fouka2019how, cao2022clans}, a setting we extend to in \sec~\ref{sec:ext_cond}. 
\end{remark}

\begin{remark}
Assumption~\ref{assm:fpt} (factorial parallel trends), together with Assumption \ref{assm:na} (no anticipation), ensures that $\tatt = \E[ \yia (1,1) - \yia (1,0)\mid G_{i} = 0 ] = \E[ \yia (1,1) -  \yia (1,0)]$ , which is the conditional average causal effect of exposure to the event with $g = 1$. However, Assumption~\ref{assm:fpt} does not imply that $\tatt = \E[\yia (G_{i},1) - \yia (G_{i},0)]$, the average treatment effect of exposure for all units. See Section A2.1 in the Supplementary Materials for more discussion. 
\end{remark}

Assumption~\ref{assm:er_g} below introduces an alternative exclusion restriction, which requires that in the absence of the event, $G$ would not have affected the average post-period outcome. It implies $\tc = \tgzo$ so that $\tdid$ also identifies $\tgzo$, as formalized in \prop~\ref{prop:tgz}. 

\begin{assumption}[Exclusion restriction absent the event]\label{assm:er_g}
$\E[\yia(1,0)] = \E[\yia(0,0)]$.
\end{assumption}

\begin{proposition}\label{prop:tgz}
(i) If Assumption~\ref{assm:er_g} holds, then $\tc = \tgzo$. 
(ii) If \assmott\ and \ref{assm:fpt}--\ref{assm:er_g} hold, then $\tdid = \tc = \tgzo$.  
\end{proposition}

Propositions~\ref{prop:tau_em}--\ref{prop:tgz} complete the roadmap in Figure~\ref{fig:estimands}. \ddef~\ref{def:FDID2} below formalizes the {\it FDID research design} as the combination of the FDID setting in \ddef~\ref{def:FDID_setting} with these identification results.

\begin{definition}\label{def:FDID2}
The {\it FDID research design} consists of (i) the FDID setting in \ddef~\ref{def:FDID_setting}, including \assmaef; 
(ii) the identification results in Propositions \ref{prop:tau_em}--\ref{prop:tgz}, under which $\tdid$ identifies 
 (a) $\te$ given Assumptions \ref{assm:ae}--\ref{assm:pta}, 
 (b) $\tatt$ given Assumptions \ref{assm:ae}--\ref{assm:er}, 
 (c) $\tc$ given \assmc, and 
 (d) $\tgzo$ given Assumptions \ref{assm:ae}--\ref{assm:pta} and \ref{assm:fpt}--\ref{assm:er_g}.
\end{definition}

\FloatBarrier

\section{Extension to Conditionally Valid Assumptions}\label{sec:ext_cond}

In many applications, the canonical and factorial parallel trends assumptions in Assumptions \ref{assm:pta} and \ref{assm:fpt} are plausible only after conditioning on baseline covariates $\xxi$ in addition to $\gi$. In this section, we present the corresponding identification and estimation results and connect them to regression methods commonly used in applied research. The unconditional setting is a special case where $\xxi = \emptyset$.

The main takeaways are twofold. First, all results in Section~\ref{sec:identification} extend to the conditional setting. Under suitable assumptions, the conditional DID estimand identifies the conditional effect modification and \cmf, and averaging over covariates yields the marginal effects. Second, coherent with Remark~\ref{rmk:cross-sec}, standard cross-sectional methods based on unconfoundedness \citep{rosenbaum1983central}, such as outcome regression and inverse propensity score weighting, carry over to the FDID setting when $(\gi, \dyi, X)_{i=1}^n$ are treated as the data. These approaches resemble the covariate-adjustment methods used in canonical DID under the conditional parallel trends assumption \citep[e.g.,][]{roth2023s}. We focus here on stratification and outcome regression using linear and TWFE specifications, and provide details on inverse propensity score weighting in Section A2.6 in the Supplementary Materials.

\subsection{Identification}\label{sec:x_iden}

Let $\xxi$ denote the vector of covariates beyond $\gi$, taking values in $\mx \subseteq \mathbb R^p$. Assumption \ref{assm:overlap} states the overlap condition that ensures the conditional expectation $\E[\ \cdot \mid \gig, \xxi]$ is well defined for $g = 0,1$. We maintain this assumption throughout the rest of the paper. 

\begin{assumption}[Overlap] \label{assm:overlap}
For all $\ximx$, $\pr( \gio \mid \xix)\in (0,1)$.
\end{assumption}

Under \assmol, define
\begin{equation}\label{eq:tau_X}
\begin{array}{lll}
\tdid(x) &=& \E[\dyi \mid \gio , \xix]- \E[\dyi \mid \giz, \xix], \\
\te(x) &=& \E[\tizgo  \mid \gio , \xix] - \E[\tizgz  \mid \giz, \xix],\\
\tc(x) &=& \E[\tic \mid \xix] = \E[\tizgo  - \tizgz  \mid \xix]
\end{array}
\end{equation}
as the {\it conditional} DID estimand, effect modification, and causal moderation, respectively, generalizing $(\tdid, \tem, \tc)$ in \eqref{eq:tdid} and \ddef~\ref{def:estimands1}. 
Define their marginal counterparts as
\begin{equation}\label{eq:tdx}
\tdx = \E[\tdxx ], \quad \tex = \E[\tem(\xxi)], \quad\tc = \E[\tc(\xxi)],
\end{equation}
where the expectations are taken over the marginal distribution of $\xxi$. 
Since $\tdid(X_i)$ and $\te(X_i)$ compare $\dyi$ and $\tizg$, respectively, between the two levels of $\gi$ conditional on $X_i$, their marginal averages $(\tdx, \tex)$ in \eqref{eq:tdx} generally differ from $(\tdid,\tem)$ unless $\gi$ and $\xxi$ are independent.
Assumptions~\ref{assm:cpt}--\ref{assm:cfpt} below build on \assmol, and extend the canonical and factorial parallel trends assumptions to the conditional setting.

\begin{assumption}[Conditional canonical parallel trends]\label{assm:cpt}
For all $\ximx$, 
$\E[\dyi(\gi, 0) \mid \gio , \xix] = \E[\dyi(\gi, 0) \mid \giz, \xix]$. 
\end{assumption}

\begin{assumption}[Conditional factorial parallel trends]\label{assm:cfpt}
For all $\ximx$, 
$\E[\dyi(g, z) \mid \gio , \xix] = \E[\dyi(g, z) \mid \giz, \xix]$ for $g, z= 0,1$.
\end{assumption}
\noindent Echoing the discussion below Assumption~\ref{assm:fpt}, a sufficient condition for Assumption \ref{assm:cfpt} is $G_i \indep \{\dyi(g,z):g, z = 0,1\} \mid \xxi$, which states that $\gi$ is as-if randomly assigned with respect to changes in all four potential outcomes, conditional on covariates $\xxi$. This condition is analogous to the unconfoundedness assumption for the cross-sectional data $(\dyi, \gi, \zi, \xxi)$. Like unconfoundedness, this assumption is inherently untestable; researchers can only assess its plausibility through auxiliary checks, such as placebo or sensitivity analyses \citep{imbens2024comparing}.

Corollary \ref{cor:did_X} extends Propositions~\ref{prop:tau_em} and \ref{prop:tauc}, and establishes the identification of $\tex$ and $\tc$.

\begin{corollary}\label{cor:did_X} 
(i) If Assumptions \ref{assm:ae}--\ref{assm:na} and \ref{assm:overlap}--\ref{assm:cpt} hold, then $\tdid(x) = \te(x)$ and $ \tdx=\tex$.
(ii) If Assumptions \ref{assm:ae}--\ref{assm:na} and \ref{assm:overlap}--\ref{assm:cfpt} hold, then $\tdid(x)= \te(x) = \tc(x)$ and $ \tdx =\tex = \tc$. 
\end{corollary}
Corollary \ref{cor:did_X}(i) follows from \prop~\ref{prop:tau_em} and ensures that $\tdx$ identifies $\tex$ under {\assmcpt}.
Corollary \ref{cor:did_X}(ii) follows from \prop~\ref{prop:tauc} and ensures that $\tdx$ identifies $\tex = \tc$ under {\assmccfpt}.
Moreover, Corollary \ref{cor:did_X} shows that $\tdid(x)$ allows us to identify the expected values of $\tem(\xxi)$ and $\tc(\xxi)$ 
under any distribution of $\xxi$, not just the marginal one. In particular, we can recover group-specific expectations, $\E[\te(X_i) \mid G_i = g]$ and $\E[\tc(X_i) \mid G_i = g]$, by averaging $\tdxx$ over the conditional distribution of $\xxi$ given $\gig$, paralleling the ATT and group causal moderation $\E[\tic \mid \gig]$ in Remark \ref{rmk:att}.

\subsection{Estimation\label{sec:est_panel}}
To apply Corollary \ref{cor:did_X}, we need to estimate $\tdxf$. From \eqref{eq:tau_X}, it suffices to estimate $\edygx$. Stratification and outcome regression are two approaches, suited to categorical and continuous $\xxi$, respectively.

For categorical $\xxi$ with $K$ levels indexed by $k = 1, \dots, K$, 
we can estimate $\E[\dyi \mid \gi=g, \xxi=k]$ by the sample average of $\dyi$ among units with $(\gi,\xxi)=(g,k)$, denoted by $\hdy(g,k)$. The resulting estimator of $\tdid(k)$ is $
\htd(k) = \hdy(1, k)-\hdy(0, k)$, as
the {\it stratum-specific DID estimator} based on units with $\xxi = k$. The marginal estimand $\tdx$ can be estimated following \eqref{eq:tdx} as $\htdx = {n}^{-1}\sumi \htd(\xxi) = \sum_{k=1}^K \pi_k \htd(k)$, 
where $\pi_k$ is the sample proportion of units with $\xxi = k$. 
This approach also applies when $\xxi$ can be meaningfully discretized. 

For continuous $\xxi$, we can estimate $\E[\dyi \mid \gi=g, \xxi=x]$ using regression, denoted by $\hdy(g, x)$, and obtain $\tdid(x)$ and $\tdx$ from \eqref{eq:tau_X}--\eqref{eq:tdx} as
\begin{equation}\label{eq:did_om}
\htd(x) = \hdy(1, x)-\hdy(0, x), \quad \htdx = {n}^{-1}\sumi \htd(\xxi).
\end{equation}
Note that 
\cls{
\edygx = \E[\yia \mid \ggxx] - \E[\yib \mid \ggxx],
} where the two sides correspond to two common regression-based approaches to DID analysis. The left-hand side corresponds to cross-sectional linear regression of $\dyi$ on $(\gi, \xxi)$, which directly estimates $\edygx$. The right-hand side corresponds to TWFE regression of $\yit$ on $(\gi, \xxi)$ with unit and time fixed effects using {\it long-format} data, where each row corresponds to a unit-time observation. We discuss both approaches below.

\subsubsection{Linear regression of $\dyi$}\label{sec:ext_cond_reg}
\ddef~\ref{def:ols} below presents two common specifications for estimating $\E[\dyi \mid \gi=g, \xxi=x]$ using cross-sectional regression of $\dyi$.

\begin{definition}\label{def:ols}  
(i)  Let \olsi\  be the ordinary least squares (OLS) fit of the model\\
\centerline{$\dyi = \bo + \bg\gi + \bxt\xxi + \bgxt\gi \xxi  + \epi,$}
and denote the estimated coefficients by $(\hboi , \hbgi, \hbxi, \hbgxi)$.\\  
(ii) Let \olsa\ be the \lsf\ of the model\\ 
\centerline{$\dyi = \bo + \bg\gi + \bxt\xxi +\epi,$}
and denote the estimated coefficients by $(\hboa , \hbga , \hbxa)$.
\end{definition}
\olsa\ is a restricted version of \olsi\ excluding the interactions between $\gi$ and $\xxi$. 
Define\\
\centerline{
$\hdyigx = \hboi + \hbgi g + \hbxit  x + \hbgxit gx, \quad
\hdyagx = \hboa + \hbga    g + \hbxa^{\T} x$}
\noindent as the estimators of $\E[\dyi\mid \gi=g, \xxi=x]$ under {\olsi} and {\olsa}, respectively. 
Following \eqref{eq:did_om}, the DID estimators based on {\olsi} and {\olsa} are: 
\begin{align}
\text{\olsi}:\quad \htdi(x) &= \hdyi(1,x) - \hdyi(0,x) = \hbgi + \hbgxit x, \label{eq:htd_*}\\
\htdxi &= \ds\meani \htdixx = \hbgi + \hbgxit \xb, \nonumber
\end{align} 
and
\begin{align}
\text{\olsa}:\quad \htdax &= \hdya(1,x) - \hdya(0,x) = \hbga, \label{eq:htd_+}\\
\htdxa &= \ds\meani \htdaxx = \hbga, \nonumber
\end{align} 
respectively, where $\xb = \meani \xxi$ is the sample mean of $\xxi$. \prop~\ref{prop:ols} below establishes the consistency of these estimators for $\tdxf$ and $\tdx$ when the corresponding models in \ddef~\ref{def:ols} are correctly specified. 

\begin{proposition}\label{prop:ols}
(i) If $\E[\dyi \mid \gi, \xxi] = \bo+\bg\gi +\bxt\xxi+\bgxt\gi\xxi$ for some constant $(\bo, \bg, \bx, \bgx)$, then (a) $\tdxf = \bg + \bgxt x$;  $\tdx = \bg + \bgxt\E[\xxi]$ and (b) $\htdix$ and $\htdxi$ in \eqref{eq:htd_*} based on {\olsi} are consistent for $\tdid(x)$ and $\tdx$.\\ 
(ii) If $\E[\dyi \mid \gi, \xxi] = \bo+\bg\gi +\bxt\xxi$ for some constant $(\bo, \bg, \bx)$, then (a) $\tdid(x) = \tdx = \bg$ and (b) $\htdax=\htdxa = \hbga$ in \eqref{eq:htd_+} based on \olsa\ are consistent for $\tdid(x)=\tdx$.
\end{proposition}

\prop~\ref{prop:ols} justifies the use of {\olsi} and \olsa\ for estimating $\tdid(x)$ and $\tdx$ under linearity assumptions on $\E[\dyi\mid \gi, \xxi]$. The identification of $\{\tem(x), \tex\}$ under Assumption~\ref{assm:cpt}, and of $\{\tc(x), \tc\}$ under Assumptions~\ref{assm:cpt}--\ref{assm:cfpt}, then follows from Corollary \ref{cor:did_X}. Together, \prop~\ref{prop:ols} and Corollary \ref{cor:did_X} justify the use of \olsa\ and \olsi\ for DID analysis of FDID when linearity holds. 
In particular, \prop~\ref{prop:ols}(i) implies that if $\E[\xxi] = 0$, then $\bg = \tdx$, so the coefficient of $\gi$ from {\olsi}, $\hbgi$, has a direct causal interpretation as a consistent estimator of $\tex$ and $\tc$ under the corresponding assumptions. The sample version of this condition can be ensured by centering the covariates so that $\bar{X}=0$, as in \cite{hirano2001estimation} and \cite{lin2013agnostic}. To account for the uncertainty in $\hbgi$, one can implement a unit-level cluster bootstrap procedure \citep{bertrand2004much}. A subtlety is that the bootstrap samples must be generated using the original $\xxi$'s, which are then recentered within each bootstrap replication. Using pre-centered covariates for resampling yields invalid inference because it ignores the sampling variability in $\bar{X}$.

By contrast, \prop~\ref{prop:ols}(ii) shows that the more parsimonious \olsa\ may be inconsistent for $\{\tdid(x), \tdx\}$ if the effect of $\gi$ on $\dyi$ varies with $\xxi$. However, when $\E[\dyi \mid \gi, \xxi]$ is truly linear in $(\gi, \xxi)$, then $\tdid(x)=\tdx$, and the coefficient $\hbga$ from {\olsa} is consistent for their common value without requiring covariate centering. 

\subsubsection{Two-way fixed-effects regression of $\yit$}\label{sec:ext_twfe}
\ddef~\ref{def:twfe} presents two common TWFE specifications for estimating $\E[\yit \mid \gi=g, \xxi=x]$, where $t = \pre, \post$. Let $\ip$ denote an indicator for the post-period. 

\begin{definition}\label{def:twfe} 
(i) Let \twi\ be the \lsf\ of the TWFE model\\
\centerline{$\yit = b_G \gi\cdot \ip + b_X \xxi \cdot \ip + b_{GX} \gi   \xxi \cdot \ip + \alpha_{i} + \xi_{t} + \epsilon_{it},$}
where $\alpha_{i}$ and $\xi_{t}$ are unit and time fixed effects, respectively, and $\epsilon_{it}$ is an idiosyncratic error.\\ 
(ii) Let \twa\ be the \lsf\ of the TWFE model\\
\centerline{$\yit = b_{G} \gi\cdot \ip + b_{X} \xxi \cdot \ip + \alpha_{i} + \xi_{t} + \epsilon_{it}.$}
\end{definition}

\twa\ is a restricted version of {\twi} that excludes the  interactions among $\gi$, $\xxi$, and $\ip$. 
Standard OLS theory ensures that\\  
\begin{tabular}{lp{.925\textwidth}}
(i)& the estimated coefficients of $(\gi\cdot \ip,\xxi \cdot \ip, \gi   \xxi \cdot \ip)$ from \twi\ equal those of $(\gi, \xxi, \gi\xxi)$ from {\olsi};\\
(ii)& the estimated coefficients of $(\gi\cdot \ip,\xxi \cdot \ip)$ from \twa\ equal those of $(\gi, \xxi)$ from {\olsa}.
\end{tabular}\medskip\\
Thus, causal interpretation of {\twi} and \twa\ follows directly from \prop~\ref{prop:ols}. As in \olsi, when using \twi\ for FDID analysis, it is important to center the covariates so that the coefficient on $\gi \cdot \ip$ has a standalone causal interpretation.

\section{Other Extensions}\label{sec:ext_rcs_G}

In this section, we generalize our main theoretical results to accommodate repeated cross-sectional data and to allow for a general (such as discrete or continuous) baseline factor~$G$.

\subsection{Repeated Cross-Sectional Data}\label{sec:rcr}

So far, our discussion has focused on analyses based on panel data. 
In many applications, however, researchers only have access to repeated cross-sectional data, where a new set of units is sampled at each time point. We extend our framework to this setting, and unify the panel and repeated cross-sections as two sampling schemes for the same population.

\subsubsection{Unification from a sampling perspective} 
Previously, $i$ indexed observed units drawn from a common population. With slight abuse of notation, we now also let~$i$ denote a generic random unit from this population, which may or may not belong to the study sample. Our theory ensures that the conditional DID estimand $\tdxf$ identifies $\tem(x)$ and $\tc(x)$ under suitable assumptions. Panel and repeated cross-sectional data then correspond to two sampling schemes: in panel data, the same units are re-observed over time, while in repeated cross-sectional data, new units are sampled at each time. Recall from \eqref{eq:tau_X} that 
\begin{equation*}
\tdxf = \E[\dyi \mid \gio, \xix] - \E[\dyi \mid G_i = 0, \xix],
\end{equation*} 
where $\edygx = \E[\yia \mid \gig, \xix] - \E[\yib \mid \gig, \xix]$. With panel data, $\dyi$ is observed directly, so both $\edygx$ and $\eygx$ can be estimated. With repeated cross-sectional data, $\dyi$ is not observed at the unit level, so estimation proceeds via $\eygx$. In either case, stratification and TWFE regression, as discussed in \sec~\ref{sec:est_panel}, are feasible, suitable for discrete $\xxi$ and continuous $\xxi$, respectively. However, outcome regression based on the OLS specifications using $\dyi$ is only applicable in the panel setting.

\subsubsection{Nested sampling with subunits and analysis} 
A distinctive feature of many FDID applications \citep[e.g.,][]{fouka2019how} using repeated cross-sectional data is nested sampling: the baseline factor $G$ is defined at the regional level, while outcomes are measured at the individual level. For clarity, we refer to the higher-level units as regions and the lower-level (sub)units as individuals. In practice, ``individual'' may also denote other sub-regional entities such as firms, schools, households, or smaller administrative units.
Three sampling designs are common in multi-period studies under nested sampling:
\begine[(i)]
\item {\it Panel--Panel}: Both regions and individuals are sampled once and followed over time, yielding panel data at both levels. 
\item {\it Panel--Repeated Cross-section}: Regions are sampled once and followed over time, while a new set of sub-regional individuals is sampled within each region at each time point. This yields panel data at the regional level and repeated cross-sections at the individual level. 
\item {\it Repeated Cross-section--Repeated Cross-section}: A new set of regions, and hence individuals, is sampled at each time point, yielding repeated cross-sections at both levels.
\ende
When individual-level data are available, researchers can conduct analysis to estimate $\tdxf$ either at the individual level (using individual-level data) or at the regional level (by aggregating individual level data to the regional level). When individual-level data are unavailable, methods for region-level analyses also apply. 

\begine[(i)]
\item {\it Panel--Panel}: For region-level analysis, estimation reduces to the panel-data methods discussed in \sec~\ref{sec:est_panel}, treating regions as units. For individual-level analysis, the same methods apply with the unit index $i$ replaced by the subunit index $ij$ and $G_{ij} = G_i$. Details are provided in Section A2.4 in the Supplementary Materials.
\item {\it Panel--Repeated Cross-section}:  For region-level analysis, the data remain panel, so estimation again reduces to methods discussed in \sec~\ref{sec:est_panel}, treating regions as units. For individual-level analysis, stratification and TWFE regression remain valid, with standard errors clustered at the regional level. A practical complication is that covariates may lack common support across time, which we leave for future research.
\item {\it Repeated Cross-section--Repeated Cross-section}: For both region-level and individual-level analyses, stratification and TWFE regression remain applicable, with standard errors clustered at the regional level. The same covariate overlap issue arises here and remains open for future work.
\ende

In summary, FDID extends naturally to repeated cross-sectional data, with panel and repeated cross-sections viewed as alternative sampling schemes from the same population. In FDID with nested data structures, across the three common nested sampling schemes, stratification and TWFE regression remain applicable for both individual- and region-level analyses, with standard errors clustered by region.

\subsection{General Baseline Factor $G$}\label{sec:cont}
The discussion so far has assumed a binary baseline factor $G$. When $G$ takes values in a general set $\mg \subseteq \mathbb{R}$, all results from the binary case continue to hold for comparisons between any two levels $g, g' \in \mg$.

Renew $\yit(g,z)$, $\dyi(g,z)$, and $\tizg = \yia(g,1) - \yia(g,0)$ for general $\gimg$. For $g, g'\in\mg$, define
$\ticg = \tizgp -  \tizg$
as the causal moderation of $G$ when its level changes from $g$ to $g'$, generalizing $\tic$. Define
\cls{
\begin{array}{rcl}
\tdgx &=& \E[\dyi \mid \gigp, \xix]  - \E[\dyi \mid \gig, \xix],\\
\tegx 
 &=& \E[\tizgp  \mid \gig', \xix] - \E[\tizg   \mid \gig, \xix],\\
\tcgx &=& \E[\ticg \mid \xix] = \E[\tizgp -\tizg \mid \xix] \medskip
\end{array} 
}
as the conditional DID estimand, effect modification, and causal moderation when $G$ changes from $g$ to $g'$, extending $\{\tdxf, \te(x), \tc(x)\}$ in \eqref{eq:tau_X}. The marginal counterparts are $\tdxg = \E[\tdg(\xxi)]$, $\texg =\E[\teg(\xxi) ]$, and $\tcg = \E[\tcg(\xxi)] = \E[\ticg]$, extending $\tc, \tdx,\tex$ in Definitions \ref{def:estimands1} and \eqref{eq:tdx}. 

\textbf{Identification.}
All identification results in \sec~\ref{sec:x_iden} extend to general $G$. Parallel to Corollary \ref{cor:did_X}, the conditional DID estimand $\tdgx$ (i) identifies  $\tegx$ under the generalized conditional \cptf\ assumptions, and (ii) identifies $\tcgx = \tegx$ under the generalized conditional \cfptf\ assumptions. We provide the details in Section A2.5 in the Supplementary Materials. 

\textbf{Regression estimation.} 
Renew (\olsi, \olsa) and (\twi, \twa) in Definitions~\ref{def:ols}--\ref{def:twfe} with general $\gi$. The numerical equivalence between OLS and TWFE continues to hold for general $\gi$. Parallel to \prop~\ref{prop:ols}, \olsi\ and \olsa\ are consistent for estimating $\tdgx$ and $\tdxg$ when the linear models are {\correct}.

With continuous $G$, we can also study incremental effects in the sense of \citet{rothenhausler2019incremental}. For brevity, we relegate the details to Section A2.5 of the Supplementary Materials.

\section{Empirical Application}\label{sec:app}
We now reanalyze the data from our running example \citep{cao2022clans} to illustrate our theory. 
Each observation represents one county in the sample. The outcome of interest is the county-level annual mortality rate, measured as the number of deaths per thousand people. Following \citet{cao2022clans}, the baseline factor takes two forms: (i) a binary indicator for high social capital, equal to one if the per capita number of genealogy books is at or above the sample median and zero otherwise; and (ii) the logarithm of the per capita number of genealogy books plus one, a continuous measure. The Great Famine began in late 1958 and, according to most historians, ended in 1961. The set of pre-famine covariates, measured at the county level, include per capita grain production, ratio of non-farming land, urbanization ratio, distance from Beijing, distance from the provincial capital, share of ethnic minorities, suitability for rice cultivation, average years of education, and log population size. 

We apply three estimators, DID, {\olsi}  (with interactions), and {\olsa} (without interactions), with the latter two incorporate covariates. As discussed earlier, in the FDID panel setting, {\olsi} and {\olsa} are numerically equivalent to {\twfei} and {\twfea}, respectively. We use 1957, one year before the famine, as the reference year for the pre-period, with $\yib$ representing the mortality rate in county $i$ in 1957. For the post-period, we define three time windows: (i) the famine years from 1958 to 1961, (ii) the pre-famine years from 1954 to 1957, and (iii) the post-famine years from 1962 to 1966. The second time window, 1954–1957, precedes the famine and is used to construct a placebo test for the (conditional) canonical parallel trends assumption, similar to a pretrend test in canonical DID \citep{angrist2009mostly}. Within each time window, we calculate the average mortality rate for each county, which serves as $\yia$. 

\citet{cao2022clans} aim to estimate ``the effect of social capital on famine relief,'' which we interpret as the causal moderation of social capital on the effect of exposure to famine. By Proposition~\ref{prop:tau_em}, if \assmott\ (universal exposure, no anticipation, and canonical parallel trends) hold, then $\htd$ recovers effect modification, $\te$. If Assumption~\ref{assm:fpt} (factorial parallel trends) also holds, then $\htd$ recovers causal moderation, $\tc$. Similarly, with covariates, Corollary~\ref{cor:did_X} and Proposition~\ref{prop:ols} imply that under  the conditional canonical parallel trends (Assumption~\ref{assm:cpt}), $\hat\beta_{G,*}$ and $\hat\beta_{G,+}$ recover effect modification, $\tex$, under their respective model specifications.  If, in addition, the conditional factorial parallel trends (Assumption~\ref{assm:cfpt}) holds, then they identify causal moderation, $\tc$.

\begin{table}[!th]
\renewcommand{\arraystretch}{0.6}
\caption{Estimating Causal Moderation of Social Capital on the Impact of Famine}\label{table:famine}
\begin{center}
\vspace{-1em} \small
\begin{tabular}{lccc}\hline\hline
\textbf{Panel A:} Binary $G$      & DID     & {\olsi}  & {\olsa} \\  
 & ($\htd$) & ($\hat\beta_{G,*}$) & ($\hat\beta_{G,+}$) \\ \hline
Famine Years (1958--1961)     & $-2.32$           & $-2.92$  & $-2.79$     \\ 
   & $[-3.75, -0.83]$     & $[-4.29, -1.39]$    &  $[-4.12, -1.31]$  \\ \\
Pre-Famine Years (1954--1956) & $0.32$      & $0.35$      & $0.33$         \\ 
   & $[-0.10, 0.72]$     &   $[-0.04, 0.78]$   &  $[-0.06, 0.75]$      \\  \\
After Famine Ends (1962--1966) & $-0.81$    & $-0.51$        & $-0.49$     \\ 
   & $[-1.20, -0.40]$      & $[-0.90, -0.11]$       & $[-0.86, -0.11]$  \medskip   \\\hline
\textbf{Panel B:} Continuous $G$      & DID     & {\olsi}  & {\olsa} \\  
 & ($\htd$) & ($\hat\beta_{G,*}$) & ($\hat\beta_{G,+}$) \\ \hline
Famine Years (1958--1961)     & $-5.85$           & $-5.14$  & $-10.11$     \\ 
   & $[-7.74, -4.01]$     & $[-8.96, -0.57]$    & $[-13.06, -6.96]$   \\ \\
Pre-Famine Years (1954--1956) & $1.02$      & $-0.51$      & $0.69$         \\ 
   & $[-0.04, 1.99]$     & $[-2.03, 0.87]$     & $[-0.41, 1.72]$       \\  \\
After Famine Ends (1962--1966) & $-1.82$    & $-1.35$        & $-1.84$     \\ 
   & $[-2.73, -0.99]$      & $[-2.43, -0.26]$       & $[-2.67, -0.97]$    \\ \hline
\end{tabular}
\end{center}

\vspace{1em}
\spacingset{1.5}\footnotesize\textbf{Note}: In Panel A, $G$ is measured as a binary indicator of social capital, while in Panel B, $G$ is measured as a continuous variable. The first column reports DID estimates without covariates. The second and third columns report regression estimates $(\hat\beta_{G,*}, \hat\beta_{G,+})$ from {\olsi} and {\olsa}, respectively, which incorporate covariates. They equal estimates of $b_{G,*}$ and $b_{G,+}$ from {\twfei} and {\twfea}, respectively. The pre-period reference year is 1957. Brackets report 95\% bootstrap confidence intervals using the percentile method. 
\end{table}

Table~\ref{table:famine} presents the results, where Panels A and B report estimates using binary and continuous measures of $G$, respectively. 
In each panel, Column (1) reports DID estimates without covariates, while columns (2) and (3) report estimates from {\olsi} and \olsa, respectively, with covariates included.
Each row corresponds to a different comparison period relative to the reference year 1957. 
In Panel A, the first row shows that high social capital (relative to low social capital) are associated with a reduction the famine-induced increase in mortality by more than 2.9 deaths per 1,000 people per year, or nearly 12 fewer deaths per 1,000 people over the four-year period. The three estimators produce similar results and closely match those reported in the original paper using TWFE models. The estimates using the average mortality rate from 1954--1956 as $\yia$ (second row) are close to zero, which support the canonical parallel trends assumption. The estimates using the average mortality rate from 1962--1966 as $\yia$ (third row) are negative—though much smaller in magnitude than during the famine years—and statistically significant at the 5\% level, suggesting a small but lasting effect of the famine on mortality rates after it ended.

\begin{figure}[!ht]
\begin{center}
\includegraphics[width=0.7\linewidth]{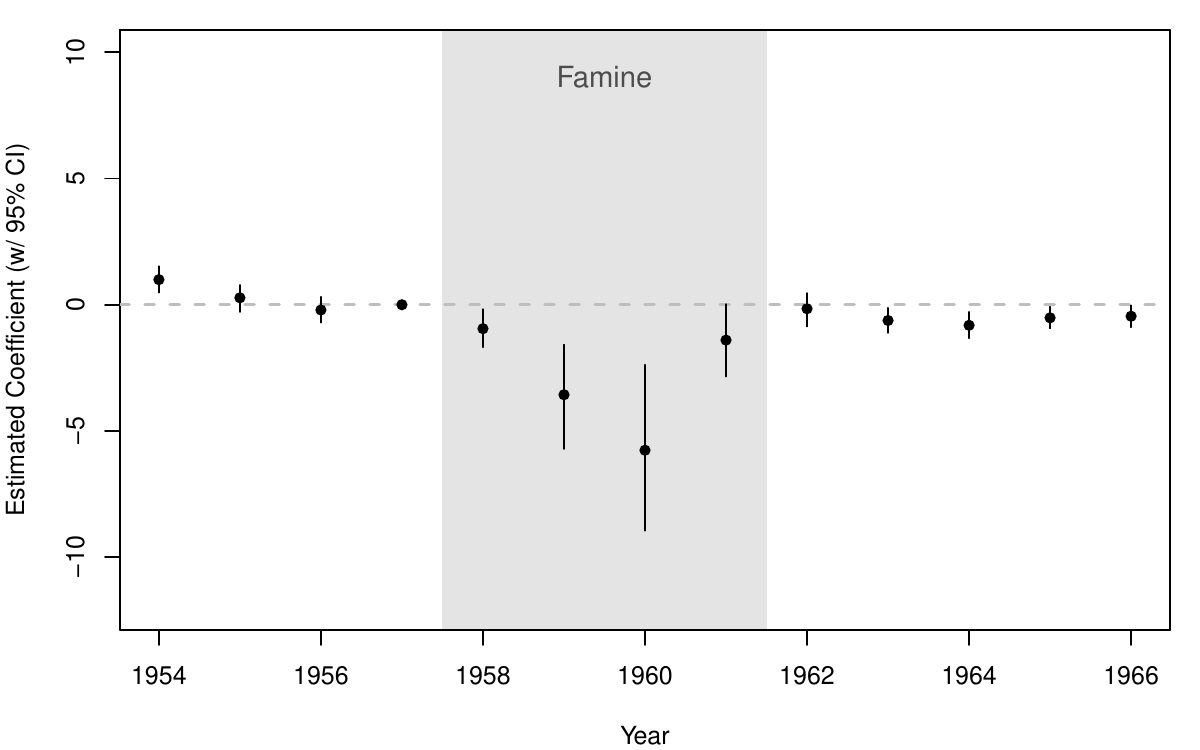} 
\end{center}
\spacingset{1.2} \caption{Estimated causal moderation over time. Each estimate is obtained from fitting {\olsi} using data in the year of interest and 1957. $G$ is a binary indicator of social capital.} \label{fig:famine.dyn}   
\end{figure}

To gain a better understanding on how the estimates evolves over time, we estimate it separately for each year from 1954 to 1966, excluding 1957 (the pre-period), defining $\yia$ as the mortality rate for each year. Figure~\ref{fig:famine.dyn} shows the results. Coherent with findings in Panel A of Table~\ref{table:famine}, the estimates are close to zero in the pre-famine years, negative during the famine (with particularly large negative estimates in 1959 and 1960), and remain negative but with much smaller magnitudes after the famine ended. 
The pre-1957 estimates are close to zero, providing suggestive evidence on the no-anticipation and (conditional) canonical parallel trends assumptions, under which the first-row estimates are effect modification. Interpreting the results as causal moderation requires the stronger (conditional) factorial parallel trends assumption, which rules out any potential confounder correlated with both social capital and the dynamics of mortality. For example, if income or governance quality are correlated with both social capital and trends in mortality, this assumption would be violated. In Section A3 in the Supplementary Materials, we present sensitivity analysis showing that a confounder with correlations to social capital and mortality changes comparable to observed covariates would have only a small impact on the estimated moderation effect. Nevertheless, we emphasize that this causal interpretation rests on a strong and inherently untestable assumption.

The patterns in Panel B mirror those in Panel A: social capital is negatively correlated with famine-induced mortality increases, but not in years leading up to the famine. For example, with $\hat\beta_{G,*}=5.15$, if linearity and constant effect assumptions hold, then a 10\% increase in the per capita number of genealogy books corresponds to a reduction of 0.5 deaths per 1,000 people per year, or about 2 fewer deaths per 1,000 people over the four-year period. Overall, the evidence supports the original authors’ claim that social capital is associated with a mitigated impact of the famine. Whether this association can be interpreted as causal, however, depends on how plausible the (conditional) factorial parallel trends assumption is.

\FloatBarrier

\section{Recommendations for Applied Researchers}\label{sec:conclude}

Based on our results, we offer several recommendations for applied researchers. First, it is important to distinguish between an estimator and a research design. An estimator is merely an algorithm applied to observed data, while a research design specifies not only the estimator but also identifying assumptions that connect it to meaningful estimands. Researchers should clearly communicate each element of the research design, especially the target estimands \citep{lundberg2021your}. In particular, the use of the DID estimator should not be conflated with the canonical DID research design. 

Second, researchers should be cautious when interpreting DID estimates as causal in FDID. Unless the exclusion restriction that exposure $Z$ has no effect on one of the two groups is plausible, the DID estimator under the no anticipation and canonical parallel trends assumptions identifies only effect modification of $G$, not the causal effects of $Z$ or $G$. To recover causal moderation by $G$,  we propose a factorial parallel trends assumption, which requires mean independence between $G_i$ and $\Delta Y_i(g,z)$ given $\xxi$.  Intuitively, it eliminates any unobserved confounders that are correlated with both $G$ and the outcome dynamics. Like unconfoundedness in the cross-sectional setting, this is a strong and untestable assumption. To identify $G$’s conditional effect given exposure, a quantity many studies aim to estimate, an additional exclusion restriction is required, stating that $G$ has no effect on the outcome when $Z = 0$. Table~\ref{tab:summary} summarizes these results and illustrates the interpretation of each estimand using the running example.

\begin{table}[!htbp]
\centering\spacingset{1.2} 
\renewcommand{\arraystretch}{1.5}
\caption{Estimands, Interpretations, and Identifying Assumptions under FDID}
\label{tab:summary}\footnotesize
\begin{tabular}{p{3.8cm}p{6.5cm}p{5cm}}
\hline\hline
\textbf{Estimand} & \textbf{Interpretation in the running example} & \textbf{Key identifying assumptions needed} \\ 
\hline
$\te$, effect modification & How did the average effect of exposure to famine on mortality differ between counties with high and low levels of social capital? & No anticipation, parallel trends \\
$\tatt$, the average effect of exposure for group $G=1$ & What was the average effect of exposure to famine on mortality in counties with high social capital? & No anticipation, parallel trends, exclusion restriction on $Z$ in group $G=1$ \\
$\tc$, causal moderation & To what extent did high (vs. low) social capital mitigate the impact of exposure to famine on mortality? & No anticipation, parallel trends, factorial parallel trends \\
$\tgzo$, $G$'s conditional effect given exposure & What was the average causal effect of social capital on mortality during the famine years? & No anticipation, parallel trends, factorial parallel trends, exclusion restriction on $G$ absent the event\\
\hline
\end{tabular}
\end{table}
\renewcommand{\arraystretch}{1}

Third, for estimation, in the absence of covariates, a TWFE regression including the interaction between $G$ and $\ip$ is numerically equivalent to the DID estimator. With additional covariates $\xxi$, standard cross-sectional methods based on unconfoundedness apply to $(\Delta Y_i,G_i,X_i)$, where $\Delta Y_i$ is the before-after difference of the outcome. Outcome regression, such as linear regression of $\Delta Y_i$ on $(1,G_i,X_i,G_iX_i)$, is commonly used when $X_i$ is continuous and is numerically equivalent to a TWFE regression with the interaction $G_iX_i \cdot \ip$. In both approaches, it is important to center $X_i$ to ensure that key regression coefficients admit a standalone causal interpretation. We discuss alternative estimation strategies in Section A2.6 of the Supplementary Materials and provide software support for their implementation through the \texttt{fdid} package in \texttt{R}.

\section*{Conflict of Interest Statement}

The authors claim no conflicts of interest.

\section*{Acknowledgment}\small

We thank Anran Liu and Rivka Lipkovitz for excellent research assistance. Peng Ding acknowledges support from the U.S. National Science Foundation (grants \# 1945136 and \# 2514234). We are grateful to Justin Grimmer, Erin Hartman, Jens Hainmueller, Laura Hatfield, Dennis Shen, Eric Tchetgen Tchetgen, Ye Wang, and seminar participants at Berkeley, MIT, Princeton, UW–Madison, UCSD, UCLA, Yale, PolMeth 2024, ACIC 2025, and the Stanford Online Causal Inference Seminar, as well as three anonymous reviewers and the Editor Hongtu Zhu, for their valuable comments. We also thank Jairui Cao and Chuanchuan Zhang for sharing the data used in \citet{cao2022clans}.

\bibliographystyle{chicago}   
{\small
\spacingset{1.4}
\bibliography{refs_panel.bib}
}

\clearpage

{
\iftrue

\appendix
\newpage
\setcounter{equation}{0}
\setcounter{section}{0}
\setcounter{figure}{0}
\setcounter{example}{0}
\setcounter{proposition}{0}
\setcounter{corollary}{0}
\setcounter{theorem}{0}
\setcounter{table}{0}
\setcounter{condition}{0}
\setcounter{definition}{0}
\setcounter{assumption}{0}
\setcounter{lemma}{0}
\setcounter{remark}{0}

\renewcommand {\thedefinition} {A\arabic{definition}}
\renewcommand {\theassumption} {A\arabic{assumption}}
\renewcommand {\theproposition} {A\arabic{proposition}}
\renewcommand {\theexample} {A\arabic{example}}
\renewcommand {\thefigure} {A\arabic{figure}}
\renewcommand {\thetable} {A\arabic{table}}
\renewcommand {\theequation} {A\arabic{equation}}
\renewcommand {\thelemma} {A\arabic{lemma}}
\renewcommand {\thesection} {A\arabic{section}}
\renewcommand {\thetheorem} {A\arabic{theorem}}
\renewcommand {\thecorollary} {A\arabic{corollary}}
\renewcommand {\thecondition} {A\arabic{condition}}
\renewcommand {\thepage} {A\arabic{page}}
\renewcommand {\theremark} {A\arabic{remark}}

\setcounter{page}{1}
\spacingset{1.5}



\begin{center}
\bf 
{\LARGE Factorial Difference-in-Differences}\\
{\Large  Supplementary Materials}
\end{center}
\vspace{2em}

\noindent\rule{\linewidth}{1pt}
\begin{itemize}\bf
    \item[] \sec~\ref{sec:proof_main}: Proofs.  
    \item[] \sec~\ref{sec:additional_app}: Additional theoretical results.
    \item[] \sec~\ref{sec:sm.app}: Additional information on the application.
\end{itemize}
\vspace{-0.5em}\noindent\rule{\linewidth}{1pt}

\vspace{3em}
\clearpage

\section{Proofs}\label{sec:proof_main}
We occasionally use the following abbreviations to simplify the presentation: 
\begin{center}
\begin{tabular}{lll}\hline
 Assumption \ref{assm:ae} (Universal exposure) & UE \\
 Assumption \ref{assm:na} (No anticipation) & NA \\
 Assumption \ref{assm:pta} (Canonical parallel trends) & PT \\
 Assumption \ref{assm:fpt} (Factorial parallel trends) & FPT \\\hline
\end{tabular}
\end{center}

\subsection{A Lemma}
Lemma \ref{lem:app} below gives an equivalent form of $\tem$ under Assumptions \ref{assm:na} (no anticipation) and \ref{assm:pta} (parallel trends) that we will use in multiple proofs.

\begin{lemma}\label{lem:app}
Under Assumptions \ref{assm:na}--\ref{assm:pta}, we have 
\begina
\te = \E[\dyi(1, 1) \mid \gio ] - \E[\dyi(0, 1) \mid \giz  ]. 
\enda
\end{lemma}

\begin{proof}[Proof of Lemma \ref{lem:app}]
{\assmptf} ensures 
\beginy\label{eq:pt_app}
\E[ \dyi(1,0) \mid \gio  ] = \E[ \dyi(0,0) \mid \giz   ]. 
\endy
Therefore, we have 
\beginy\label{eq:a1-a0}
 \te &=& \E[ \tau_{i, Z|G_i = 1} \mid \gio ] - \E[ \tau_{i, Z|G_i = 0} \mid \giz  ] \nonumber\\
 &\overset{\eqref{eq:pt_app}}{=}& 
 \E[ \tau_{i, Z|G_i = 1} \mid \gio ] - \E[ \tau_{i, Z|G_i = 0} \mid \giz  ] \nonumber\\
&& + \ \E[\dyi(1, 0) \mid \gio ] - \E[\dyi(0, 0) \mid \giz  ] \nonumber\\
 &=& 
 \E\Big[ \tau_{i, Z|G_i = 1} + \dyi(1, 0) \mid \gio \Big] - \E\Big[ \tau_{i, Z|G_i = 0} + \dyi(0,0) \mid \giz  \Big] \nonumber\\
&=& A_1 - A_0,  
\endy
where 
\beginy\label{eq:A1_def}
A_1 = \E\Big[ \tau_{i, Z|G_i = 1} + \dyi(1, 0) \mid \gio \Big],\quad 
A_0  
= \E\Big[ \tau_{i, Z|G_i = 0} + \dyi(0,0) \mid \giz  \Big]. 
\endy
Note that 
\beginy\label{eq:A1_ss}
\tau_{i, Z|G_i = 1} + \dyi(1, 0) &=& \yia(1, 1) - \yia(1, 0) + \yia(1, 0) - \yib(1, 0)\nonumber\\
&=& \yia(1, 1)   - \yib(1, 0)   
\endy
in the definition of $A_1$ in \eqref{eq:A1_def}. Under \assmna, plugging \eqref{eq:A1_ss} in the expression of $A_1$ in \eqref{eq:A1_def} ensures 
\beginy\label{eq:A1}
\begin{array}{lcl}
A_1 &\overset{\eqref{eq:A1_def}+\eqref{eq:A1_ss}}{=}&
 \E[\yia(1, 1) - \yib(1, 0) \mid \gio ] \\ 
&\overset{\text{Assm. \ref{assm:na} (NA)}}{=}&  \E[\yia(1, 1) - \yib(1, 1) \mid \gio ]\\ 
&=& \E[\dyi(1,1) \mid \gio  ],\smallskip\\
A_0 &\overset{\text{by symmetry}}{=}& \E[\dyi(0,1) \mid \giz  ].
\end{array} 
\endy
Plugging \eqref{eq:A1}  in \eqref{eq:a1-a0} completes the proof. 
\end{proof}

\subsection{Proofs of the results in \sec~\ref{sec:identification}}

\begin{proof}[Proof of Proposition~\ref{prop:tau_em}] 
{\assmae} ensures 
\begina
\dyi = \left\{
\begin{array}{ll}
\dyi(1,1) & \text{if $\gio $}\\
\dyi(0,1) & \text{if $\giz  $}
\end{array} \right.
\enda
such that 
\begina
\tdid &=&   \E[\dyi \mid \gio  ] - \ee[\dyi \mid \giz   ] \\
&=&  \E[\dyi(1,1) \mid \gio ] - \E[\dyi(0,1) \mid \giz  ].
\enda
In addition, Lemma \ref{lem:app} ensures 
$$
\tem =  \E [\dyi(1,1) \mid \gio  ] - \E [\dyi(0,1) \mid \giz  ]
$$ 
under Assumptions \ref{assm:na} (no anticipation) and \ref{assm:pta} (canonical parallel trends). 
These two results together ensure $\tdid = \tem$ under Assumptions \ref{assm:ae}--\ref{assm:pta}. 
\end{proof}

\begin{proof}[Proof of \prop~\ref{prop:tauc}]
We verify below 
$ \te = \tc$ under Assumptions \ref{assm:na}--\ref{assm:pta} and \ref{assm:fpt}. Their identification by $\tdid$ 
then follows from \prop~\ref{prop:tau_em}. 

First, Assumptions \ref{assm:pta}--\ref{assm:fpt} (canonical and factorial parallel trends)  together ensure
\beginy\label{eq:tauac_ss}
&&\E[ \yia(1,0) - \yib(1,0) ] - \E [ \yia(0,0) - \yib(0,0) ] \nonumber\\ &=& \E[ \dyi(1,0) ] -  \E[ \dyi(0,0) ] \nonumber\\
&\overset{\text{Assm. \ref{assm:fpt} (FPT)}}{=}& \E\left[ \dyi(1,0) \mid \gio \right] - \E\left[ \dyi(0,0) \mid \giz  \right]\nonumber\\
&\overset{\text{Assm. \ref{assm:pta} (PT)}}{=}& 0.
\endy
Equation~\eqref{eq:tauac_ss} implies 
\beginy\label{eq:fpt}
\E[ \yia(1,0) - \yia(0,0)] = \E[ \yib(1,0) - \yib(0,0)]
\endy
and, together with Lemma \ref{lem:app}, ensures
 \begina
 \te
&\overset{\text{Lemma \ref{lem:app}}}{=}& \E[\dyi(1, 1) \mid \gio ] - \E[\dyi(0, 1) \mid \giz  ] \\
&\overset{\text{Assm. \ref{assm:fpt} (FPT)}}{=}&  \E[\dyi(1, 1) ] - \E[\dyi(0, 1) ] \\
&=& 
 \E[\yia(1, 1) - \yib(1, 1)] - \E[ \yia(0, 1)- \yib(0, 1)] \\
&=& \E[\yia(1, 1) - \yia(0, 1) ] - \E[ \yib(1, 1) - \yib(0, 1)] \\
&\overset{\text{Assm. \ref{assm:na} (NA)}}{=}& \E[\yia(1, 1) - \yia(0, 1) ] - \E[ \yib(1, 0) - \yib(0, 0)] \\
&\overset{\eqref{eq:fpt}}{=}& \E[\yia(1, 1) - \yia(0, 1) ] - \E[ \yia(1, 0) - \yia(0, 0)] \\
&=& \tc. 
\enda

\end{proof}

\clearpage

\section{Additional Theoretical Results}\label{sec:additional_app}

\subsection{ATE Analogs under Assumption~\ref{assm:fpt}}

In the paper, we define $\tatt$, the ATT analog under FDID, as
\begin{eqnarray*}
\tatt &=& \E[\tizgg  \mid \gio ] \\
&=& \E[ \yia (1,1) - \yia (1,0)\mid \gio ].
\end{eqnarray*}
Lemma~\ref{lemma:ate} shows its relationship with the conditional average causal effect under the no anticipation and factorial parallel trends assumptions. 

\begin{lemma}\label{lemma:ate}
    Under Assumptions~\ref{assm:na} and~\ref{assm:fpt}, we have
\begin{eqnarray*}
    \tatt &=& \E[\yia (1,1) - \yia (1,0)\mid G_{i} = 0 ] \\
    &=& \E[ \yia (1,1) - \yia(1,0)] . 
\end{eqnarray*}
\end{lemma}

\begin{proof}
We have 
\begina
\tatt & \overset{\text{Def.}}{=} &\E[\yia (1,1) - \yia(1,0)\mid \gio ]\\
&=& \E[\yia(1,1) - \yib (1,1)\mid \gio ]  + \E[ \yib (1,1)- \yia (1,0)\mid \gio ]\\
&\overset{\text{NA}}{=}& \E[ \yia (1,1) - \yib (1,1) \mid \gio ] + \E[\yib (1,0)- \yia (1,0) \mid\gio ]\\
&\overset{\text{FPT}}{=}& \E[ \yia (1,1) - \yib (1,1) \mid \giz ] + \E[ \yib (1,0)- \yia (1,0) \mid\giz ]\\
&\overset{\text{NA}}{=}& \E[ \yia (1,1) - \yib (1,1) \mid \giz ] + \E[ \yib (1,1)- \yia (1,0) \mid\giz ]\\
&=& \E[ \yia (1,1) - \yia (1,0) \mid\giz ],
\enda 
which is the first identity. 
Therefore, we also have 
\[
\tatt 
= \E[ \yia (1,1) - \yia (1,0)],
\]
which is the second identity. 
\end{proof}

However, in the paper, we did not state that Assumption \ref{assm:fpt} implies $\text{ATT} = \text{ATE}$ to avoid confusion because some readers might interpret $\E[ \yia (G_{i},1) - \yia (G_{i},0)]$ (the average treatment effect of exposure for all units, regardless of their realized value of $G_i$) as the ATE. But
\[
\tatt \neq \E[\yia (G_{i},1) - \yia(G_{i},0)],
\]
because 
\begina
\E[\yia (G_{i},1) - \yia(G_{i},0)] 
&=& \E[\yia(0,1) -\yia(0,0) \mid \giz] \cdot \pr(\giz)\\
&&+ \E[\yia(1,1) - \yia(1,0) \mid \gio] \cdot \pr(\gio).
\enda
We can construct a simple contraction, for example, by assuming Assumption \ref{assm:er} (the exclusion restriction) holds. Under the exclusion restriction, $\E[\yia (0,1) - \yia (0,0) \mid \giz] = 0$, then
\begina
    \E[ \yia (G_{i},1) - \yia (G_{i},0)] 
&=&  \E[ \yia (1,1) - \yia (1,0) \mid \gio] \cdot \pr(\gio)\\
&=& \E[ \yia (1,1) - \yia (1,0)] \cdot \pr(\gio).\\
&=& \tatt \cdot \pr(\gio) \neq  \tatt . 
\enda
Therefore, we avoid stating that, under Assumption \ref{assm:fpt}, $\text{ATT} = \text{ATE}$ (when the latter is not clearly defined).

\subsection{Details about $\tau_{\text{\scriptsize DID-X}}$ and $\tau_{\text{\scriptsize em-X}}$}
We show in the following that generally, 
\begine[(i)]
\item\label{statement:did} $\tdid$ is not a weighted average of $\tdxx $ over any distribution of $\xxi$; 
\item\label{statement:em} $\tem$ is not a weighted average of $\tem(\xxi)$ over any distribution of $\xxi$
\ende 
unless $\xxi$ and $\gi$ are independent. 

First, it follows from $
\E[\dyi \mid \gio , \xxi] = \tdxx  + \E[ \dyi \mid \giz, \xxi]$
and the law of iterated expectations that 
\beginy\label{eq:not average_did}
\tdid &=& \E[\dyi \mid \gio ] - \E[\dyi \mid \giz  ]\nonumber\\
&=& \E\Big\{\E[\dyi \mid \gio , \xxi] \mid \gio \Big\} - \E\Big\{\E[\dyi \mid \giz, \xxi] \mid \giz  \Big\}\nonumber\\
&=& \E[\tdxx  \mid \gio ]\nonumber\\
&&+
\E\Big\{\E[\dyi \mid \giz, \xxi] \mid \gio \Big\}
- \E\Big\{\E[\dyi \mid \giz, \xxi] \mid \giz  \Big\},
\endy
where the outer expectation of the first two terms in \eqref{eq:not average_did} are with respective to the conditional distribution of $\xxi$ given $\gio $ and the outer expectation of the third term  in \eqref{eq:not average_did} is with respective to the conditional distribution of $\xxi$ given $\giz  $.
This implies Statement \eqref{statement:did}.

In addition, replacing $\dyi$ by $\tizgg$ in \eqref{eq:not average_did} ensures 
\begina
\tem &=& \E[\tizgg \mid \gio ] - \E[\tizgg \mid \giz  ]\nonumber\\
&=& \E\Big\{\E[\tizgg\mid \gio , \xxi] \mid \gio \Big\} - \E\Big\{\E[\tizgg \mid \giz, \xxi] \mid \giz  \Big\}\nonumber\\
&=& \E[\tem(\xxi) \mid \gio ]\nonumber\\
&&+
\E\Big\{\E[\tizgg \mid \giz, \xxi] \mid \gio \Big\}
- \E\Big\{\E[\tizgg\mid \giz, \xxi] \mid \giz  \Big\}.
\enda
This implies Statement \eqref{statement:em}.

\clearpage

\subsection{Reconciling FDID with Canonical DID}

Figure~\ref{fig:id2} illustrates \prop~\ref{prop:recon}(ii) and reconciles FDID with canonical DID. When Assumption~\ref{assm:er} holds, that is, when $\E[\yia(0,1) \mid \giz] = \E[\yia(0,0) \mid \giz]$, the FDID setting reduces to canonical DID.

\begin{figure}[!ht]
\begin{center} 
 \includegraphics[width = 0.65\textwidth]{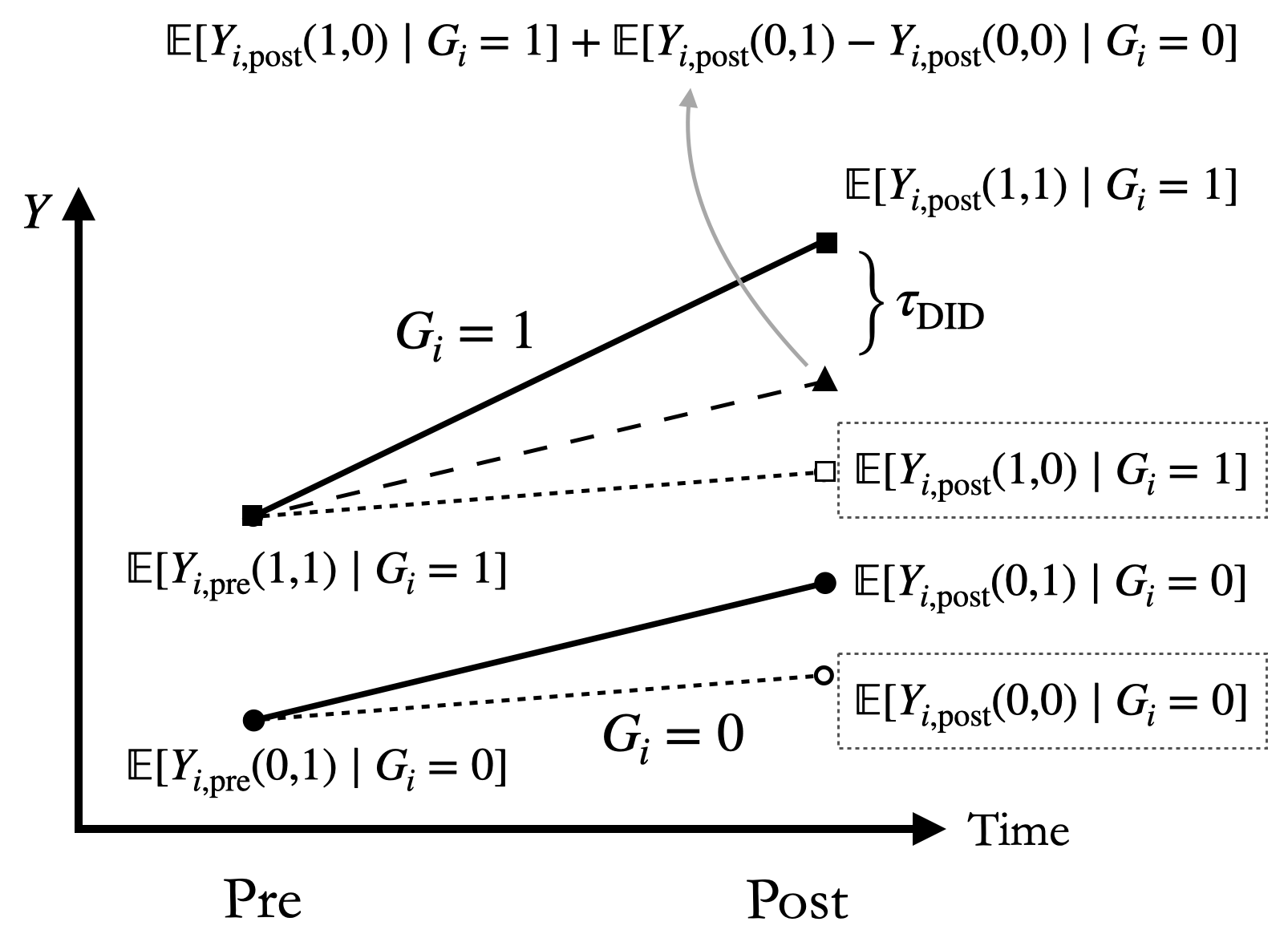} 
 \end{center}  
\spacingset{1.2}
\caption{\label{fig:id2}Reconcile FDID with Canonical DID.} \medskip
{\footnotesize \textbf{Note.} We use squares $\blacksquare$ to represent the group with $G_i = 1$, and circles $\text{\ding{108}}$ to represent the group with $G_i = 0$. {\assmna} implies that in the pre-event period, $\blacksquare = \E[\yib(1,0) \mid G_i = 1] = \E[\yib \mid G_{i} = 1]$ and $\text{\ding{108}} = \E[\yib(0,0) \mid G_i = 0] = \E[\yib \mid G_{i} = 0]$, so that $\blacksquare$ and \ding{108} represent the before and after observed outcomes of the two groups.  The short dashed line parallels the solid line for $G_i = 0$ by construction of the DID estimator, so that in the post-period, the distance between $\blacksquare$ and \ding{115} equals $\tdid$. Under \assmpt, we have \ding{115} equals $ \E[\yia(1,0)\mid G_i = 1]$ such that the distance between the post-period $\blacksquare = \E[\yia(1,1)\mid G_i=1]$ and \ding{115} also equals the ATT = $\E[\yia(1,1)-\yia(1,0)\mid G_i=1]$. Under {\assmer} (exclusion restriction), the post-period \ding{108} = $\E[\yia(0,1) \mid G_i = 0]$ and $\text{\mcirc} = \E(\yia(0,0)\mid G_i = 0]$ coincide, and only \ding{108} is shown. This ensures $\square = \E[\yia(1,0)\mid G_i = 1]$ (not shown) merges into \ding{115}, so Figure \ref{fig:id1} reduces to Figure \ref{fig:id2}. This illustrates the correspondence between FDID and canonical DID under {\assmer}.}
\end{figure}

\clearpage

\subsection{Details for \sula\ in \sec~\ref{sec:rcr}}

In this subsection, we discuss the identification and estimation of FDID using repeated cross-sectional (RCS) data from a nested sampling perspective. We consider three scenarios described in the main text. The preferred estimation strategies are summarized in Table~\ref{tab:rcs}.

\begin{table}[!ht]
    \centering
\begin{center}
    \caption{FDID with RCS Data: A Nested Sampling Perspctive}
    \label{tab:rcs}
\begin{tabular}{l|c|c}\hline
& Unit-level analysis & Subunit-level analysis\\\hline
\ppp & \multirow{2}{*}{\begin{tabular}{c}Stratification + OLS or TWFE \\ (\sec~\ref{sec:est_panel})\end{tabular}} & Stratification + OLS or TWFE \\\cline{1-1}\cline{3-3}
\ppr & & \multirow{2}{*}{Stratification + TWFE} \\\cline{1-2}
\rr & Stratification + TWFE & \\\hline
\end{tabular}
\end{center}
\end{table}

\subsubsection{DID estimand at the \sul}\label{sec:su_app_population}
Assume a {\it two-stage} sampling mechanism, also known as {\it grouped} or {\it cluster} sampling, where  
\begine[(i)] 
\item the first stage selects a simple random sample of units from a population of units;
\item the second stage selects a simple random sample of subunits\ within each selected unit. 
\ende
Let $i$ index a generic random unit from the population, and let $ij$ index a generic random \su\ within unit $i$.
Let $\gi \in \{0,1\}$ denote the baseline factor $G$ of unit $i$.
Let $\yijpre, \yijpost \in \mbr$, $\gij \in \{0,1\}$, and $\xij\in \mbr^p$ denote the before and after outcomes, baseline factor $G$, and covariates for subunits\ $ij$, where $\gij = \gi$. 
The two-stage sampling mechanism ensures that $\{(\yijpre, \yijpost), \gij, \xij\}$ are identically distributed across all subunits\ in the population.
Let $\dyij = \yijpost-\yijpre$. 
The conditional DID estimand at the \sul\ is
\begina
\tds(x) &=& \E[\dyij \mid \gio, \xix] - \E[\dyij \mid \giz, \xix],
\enda
where we use the prime superscript ($^\prime$) to denote \sulh\ quantities. Given\\
\centerline{$\edyijgx = \E[\yijpost \mid \gig , \xijx] - \E[\yijpre \mid \gig , \xijx]$,} to estimate $\tdsxf$, it suffices to estimate either $\edyijgx$ or $\E[\yijt \mid \ggxxs ]$.
Depending on the definition of the \ulh\ outcome $\yit$ in terms of $\yijt$, $\tds(x)$ may or may not equal $\tdxf$.

\subsubsection{Regression estimation of $\tau_\textup{DID}'(x)$}
Let $\ist$ index the $i$-th unit in the sample at time $t$, and let $\istj$ denote the $j$-th \su\ in unit $\ist$.
With slight abuse of notation, let $\git$ denote the level of the baseline factor $G$ of unit $\itp$, and let $(\yitjt, \xitj)$ denote the outcome and covariates of \su\ $\itj$, with $\yitjt = Y_\itjt$ under the population notation in \sec~\ref{sec:su_app_population}. 
This notation covers the \pp, \prf, and \rr\ schemes. 
Under the \pps, we can suppress the subscript $t$ in $\itp$, and use $i$ and $ij$ to index the units and subunits, respectively, with $\{(\git, \yitjt, \xitj): t = \pre, \post\}$ reduced to $(\gi, \yijpre, \yijpost, \xij)$.

\paragraph{TWFE for panel or repeated cross-section \suld.}
Parallel to Definition~\ref{def:twfe}, two common specifications for \sulh\ TWFE regression based on $\{(\git, \yitjt, \xitj): t = \pre, \post\}$ are the OLS fits of the following TWFE models, with and without the three-way interactions among $\git$, $\xijt$ and $\ip$: 
\beginy\label{eq:twfe_ij}
\begin{array}{ll}
\text{\twis}:&\yitjt = b_G \git\cdot \ip + b_X \xitj \cdot \ip + b_{GX} \git   \xitj \cdot \ip + \ait + \xi_{t} + \epsilon_{\ijt},\\
\text{\twas}:&\yitjt = b_G \git\cdot \ip + b_X \xitj \cdot \ip + \ait + \xi_t + \epsilon_{\ijt},
\end{array}
\endy
where $(\ait, \ist, \epsilon_\ijt)$ denote the fixed effect of unit $\ist$, the time fixed effect, and an idiosyncratic error of \su-time pair $(\istj,t)$. 
With panel \suld, \eqref{eq:twfe_ij} simplifies to 
\begina
\begin{array}{ll}
\text{\twis}:&\yijt = b_G \gi\cdot \ip + b_X \xij \cdot \ip + b_{GX} \gi  \xij \cdot \ip + \ait + \xi_{t} + \epsilon_{\ijt},\\
\text{\twas}:&\yijt = b_G \gi\cdot \ip + b_X \xij \cdot \ip + \ait + \xi_t + \epsilon_{\ijt}. 
\end{array}
\enda

\paragraph{OLS for panel \suld.}  
With panel \suld\ $(\gi, \yijpre, \yijpost, \xij)$, parallel to Definition~\ref{def:ols}, two common specifications for \sulh\ cross-sectional regression based on $(\dyij, \gi, \xij)$ are the OLS fits of the following  models, with and without interactions between $\gi$ and $\xij$: 
\begina
\text{\olsis}:&&\dyij = \bo + \bg\gi + \bxt\xij + \bgxt\gi \xij  + \epij,\\
\text{\olsas}:&&\dyij = \bo + \bg\gi + \bxt\xij  + \epij.
\enda

\paragraph{Justification.}
All results in \sec~\ref{sec:est_panel} at the unit level extend to the \sul: 
\begine[(i)]
\item \olsis\ and \twis\ are numerically equivalent, and are consistent for estimating $\tds$ if $\edyijgx$ is linear in $(\gi, \xij, \gi\xij)$.
\item \olsas\ and \twas\ are numerically equivalent, and are consistent for estimating $\tds$ if $\edyijgx$ is linear in $(\gi, \xij)$.
\ende  
In practice, \sulh\ outcomes may depend on unit size. When this is the case, it is important to include unit size as a covariate in $\xijt$ to improve the plausibility of the linearity assumption.

\subsection{Details for extension to general baseline factor $G$}

This subsection discusses results when $G$ is extended to be a discrete or continuous variable.

\subsubsection{Identification}
Assumption \ref{assm:G} and Corollary \ref{cor:G_did} below generalize Corollary \ref{cor:did_X} to general $G$, clarifying the conditions for  $\texg$ and $\tcg$ by $\tdxg$.  
\begin{assumption}\label{assm:G} 
\begine[(i)]
\item\label{item:G_na} {\it No anticipation:} $\yib(g, 0) = \yib(g,1)$ for all $\gimg$. 
\item {\it Overlap:} For all $\ximx$, 
\begini 
\item For discrete $G$, $\pr( \gig \mid \xix)\in (0,1)$ for all $\gimg$;
\item For continuous $G$, the conditional density satisfies $f_{G \mid X}( g \mid \xix) > 0$ for all $\gimg$;
\endi 
\item\label{item:G_cpt} {\it Conditional canonical parallel trends:} $\E[\dyi(\gi, 0) \mid \gig, \xxi]$ is constant across $\gimg$.
\item\label{item:G_cfpt} {\it Conditional factorial parallel trends:} For $\gimg$ and $z = 0,1$, $\E[\dyi(g, z) \mid \gig', \xxi]$ is constant across $g' \in \mg$.
\ende 
\end{assumption}

\begin{corollary}\label{cor:G_did} 
\begine[(i)] 
\item\label{item:G_cor_em} If Assumption~\ref{assm:ae} and Assumption~\ref{assm:G}\eqref{item:G_na}--\eqref{item:G_cpt} hold, then $\tdgx = \tegx$ and $ \tdxg=\texg$.
\item\label{item:G_cor_cm} If Assumption~\ref{assm:ae} and Assumption \ref{assm:G}\eqref{item:G_na}--\eqref{item:G_cfpt} hold, then 
$\tdgx= \tegx = \tcgx$ and $ \tdxg =\texg = \tcg$. 
\ende
\end{corollary}

\subsubsection{Regression estimation} 
Echoing \sec~\ref{sec:est_panel}, regression gives a convenient way to estimate $\tdgx$ and $\tdxg$ for general $G$. We establish below the identification results based on regression coefficients. 

Renew $(\hboi, \hbgi, \hbxi, \hbgxi)$ and $(\hboa, \hbga, \hbxa)$ as the coefficients from \olsi\ and \olsa, respectively, with general $\gi$. 
Renew 
\begina
\hdyigx = \hboi + \hbgi g + \hbxit  x + \hbgxit gx, \quad
\hdyagx = \hboa + \hbga g + \hbxa^{\T} x
\enda
for $(g,x) \in \mg\times \mx$
as the corresponding estimators of $\E[\dyi\mid \gi=g, \xxi=x]$,
and 
\beginy\label{eq:htd_*_G}
\begin{array}{rclcl}
\htdgi(x) &=& \hdyi(g',x) - \hdyi(g,x) &=& (g'-g)(\hbgi + \hbgxit x),\\
\htdxgi &=& \displaystyle\meani \htdgi(\xxi) &=& (g'-g)(\hbgi + \hbgxit \xb),
\end{array}\\
\label{eq:htd_+_G}
\begin{array}{rclcl}
\htdga(x) &=& \hdya(g',x) - \hdya(g,x) &=& (g'-g)\hbga,\\
\htdxga  &=& \displaystyle\meani \htdga(\xxi) &=& (g'-g)\hbga 
\end{array}
\endy
as the corresponding DID estimators, generalizing $(\htdi(x), \htdxi)$ and $(\htda(x), \htdxa)$. 
\prop~\ref{prop:ols_G} below generalizes \prop~\ref{prop:ols} to general $G$. 
Standard OLS theory ensures that the numerical equivalence between \olsa\ and \twa, and that between \olsi\ and \twi, also holds for general $G$.

\begin{proposition}\label{prop:ols_G}
\begine[(i)]
\item\label{item:ols_G_1} If $\E[\dyi \mid \gi, \xxi] = \bo+\bg\gi +\bxt\xxi+\bgxt\gi\xxi$ for some constant $(\bo, \bg, \bx, \bgx)$, then 
\begina
\tdgx = (g'-g)(\bg + \bgxt x), &&\tdxg = (g'-g)(\bg + \bgxt\E[\xxi]),
\enda 
 for  $g, g'\in\mg$ and $\ximx$, so that $\htdgix = (g'-g)(\hbgi + \hbgxit x)$ and $\htdxgi  = (g'-g)(\hbgi + \hbgxit \xb)$ in \eqref{eq:htd_*_G} based on {\olsi} are consistent for $\tdid(x)$ and $\tdx$. 
\item\label{item:ols_G_2} If $\E[\dyi \mid \gi, \xxi] = \bo+\bg\gi +\bxt\xxi$ for some constant $(\bo, \bg, \bx)$, then 
\begina
\tdgx = \tdxg = (g'-g)\bg, 
\enda
 for  $g, g'\in\mg$ and $\ximx$, so that $\htdgax=\htdxga = (g'-g)\hbga$ in \eqref{eq:htd_+_G} based on \olsa\ are consistent for $\tdgx=\tdxg$.
\ende
\end{proposition}

\begin{proof}[\bf Proof of Corollary \ref{cor:G_did}\eqref{item:G_cor_em}] 
Under Assumption \ref{assm:G}, we have 
\begina
\tdgx &=& 
\E[\dyi \mid \gigp, \xix ] - \E[\dyi \mid \gig, \xix ] \\
&\overset{\text{Assumption \ref{assm:ae}}}{=}& 
\E[\dyi(g', 1) \mid \gigp, \xix] - \E[\dyi(g, 1) \mid \gig, \xix ]\\
&\overset{\text{Assumption \ref{assm:G}\eqref{item:G_cpt}}}{=}& 
\E[\dyi(g', 1) \mid \gigp, \xix] - \E[\dyi(g', 0) \mid \gigp, \xix]\\
&& - \Big\{\E[\dyi(g, 1) \mid \gig, \xix] -\E[\dyi(g, 0) \mid \gig, \xix] \Big\}\\
&=& 
\E[\dyi(g', 1) -\dyi(g', 0) \mid \gigp, \xix] \\
&& -  \E[\dyi(g, 1) - \dyi(g, 0) \mid \gig, \xix]  \\
&\overset{\text{Assumption \ref{assm:G}\eqref{item:G_na}}}{=}& \E[\tau_{Z\mid G = g'} \mid \gigp, \xix] - \E[\tau_{Z\mid G = g} \mid \gig, \xix]\\
&=& \tegx. 
\enda
\end{proof}

\begin{proof}[\bf Proof of \prop~\ref{prop:ols_G}] 
When $\E[\dyi \mid \gi, \xxi] = \bo + \bg\gi+ \bxt\xxi + \bgxt\gi\xxi$, we have 
\begina
\tdgx &=& \E[\dyi \mid \gigp, \xix ] - \E[\dyi \mid \gig, \xix ]\\
&=&(g'-g)(\bg + \bgxt x) 
\enda
by definition, and properties of least squares ensure
\begina
\plim (\hbgi, \hbxi, \hbgxi) = (\bg, \bx, \bgx). 
\enda
This ensures \prop~\ref{prop:ols_G}\eqref{item:ols_G_1}. 

\prop~\ref{prop:ols_G}\eqref{item:ols_G_2} is a direct consequence of \prop~\ref{prop:ols_G}\eqref{item:ols_G_1} with $\bgx = 0$. 
\end{proof}

\subsubsection{Extension to the incremental effects}

When $\mg$ is an interval, define 
\begin{eqnarray*}
\delta_{\textsc{did}}(g, x) &=& \frac{\partial \E[\dyi \mid \gig, \xix] }{\partial g},\\
\delta_{\text{em}}(g, x) 
 &=& \frac{\partial \E[\tizg   \mid \gig, \xix] }{\partial g},\\
\delta_{\text{cm}}(g, x) &=& \frac{\partial  \E[ \tizg \mid \xix] }{\partial g}
\end{eqnarray*}
as the ``incremental effects'' in the sense of \citet{rothenhausler2019incremental}. 

Parallel to Corollary \ref{cor:did_X}, the DID estimand $\delta_{\textsc{did}}(g, x)$ (i) identifies  $\delta_{\text{em}}(g, x)$ under the generalized conditional \cptf\ assumptions, and (ii) identifies $\delta_{\text{em}}(g, x) = \delta_{\text{cm}}(g, x)$ under the generalized conditional \cfptf\ assumptions.

Under the assumption in Proposition \ref{prop:ols_G}(\ref{item:ols_G_1}), we have $\delta_{\textsc{did}}(g,x) = \bg + \bgxt x$, and under the assumption in Proposition \ref{prop:ols_G}(\ref{item:ols_G_2}), we have $\delta_{\textsc{did}}(g, x) = \bg$. 
In addition, 
(i) when \olsi\ is correctly specified, then $\delta_{\textsc{did}}(g,x) = \bg + \bgx x$ is constant with each level of $x$, and the estimated coefficient of $\gi$ is consistent for $\bg$; (ii) when \olsa\ is correctly specified, then $\delta_{\textsc{did}}(g,x)=\bg$, and the estimated coefficient of $\gi$ from \olsa\ is consistent for $\bg$.

\subsection{Identification by inverse propensity score weighting}
Outcome regression and inverse propensity score weighting are two leading methods in causal inference from observational cross-sectional data. 
We extend below the discussion in \sec~\ref{sec:ext_cond} to inverse propensity score weighting, and establish the corresponding identification results in FDID.

Let $e(x) = \pr( \gio \mid \xix)$ denote the propensity score of $\gi$ given $\xxi$ \citep{rosenbaum1983central}. Define
\begina
\tipw = \E\left[\frac{\gi}{e(\xxi)} - \frac{1-\gi}{1-e(\xxi)}\cdot \dyi \right]
\enda
as the inverse-propensity-score-weighted estimand by treating $\{\dyi, \gi: i = \ot{n}\}$ as the cross-sectional input data.
Define
\begina
\tipw(x)= \E\left[\frac{\gi }{e(x)} \cdot \dyi- \frac{1-\gi  }{1-e(x)}\cdot \dyi \mid \xix \right]
\enda
as the conditional version of $\tipw$ with 
$\tipw = \E [\tipw(\xxi) ]$. 
Define
\begina
\tau_{\ipw \mid G = 1} &=& \E[\dyi \mid \gio ] - \E\left[\dfrac{e(\xxi)}{e} \cdot \dfrac{1-\gi}{1-e(\xxi)} \cdot \dyi \right],\\
\tau_{\ipw \mid G = 0} &=& \E\left[\dfrac{1-e(\xxi)}{1-e} \cdot \dfrac{\gi}{e(\xxi)} \cdot \dyi \right] - \E[\dyi \mid \giz  ],
\enda
where $e = \pr(\gi=1)$.
\prop~\ref{prop:ipw} below ensures the numerical equivalence between $\tipw(x)$ and $\tdid(x)$ and that between $\tau_{\ipw\mid G = g}$ and $\E[\tdxx  \mid \gig]$ for $g = 0,1$.
Identification based on $\tipw(x)$, $\tipw$, and $\tau_{\ipw\mid G = g} \ (g = 0,1)$ then follows from Corollary \ref{cor:did_X}. 
Specifically, $\tau_{\ipw\mid G = g}$ identifies $\E[\tem(\xxi) \mid \gig]$ under Assumptions \ref{assm:ae}--\ref{assm:na} and \ref{assm:overlap}--\ref{assm:cpt} (universal exposure, no anticipation, overlap, and conditional canonical parallel trends), and identifies $\E[\tc(\xxi) \mid \gig]$ if we further assume \assmcfpt. 

\begin{proposition}\label{prop:ipw}
Under Assumption \ref{assm:overlap}, we have
\begine[(i)] 
    \item $\tipw(x) = \tdid(x)$ with $\tipw = \tdx$.
    \item $\tau_{\ipw \mid G = g} = \E[\tdxx  \mid \gig]$ for $g = 0,1$.
\ende
\end{proposition}

\begin{proof}[Proof of \prop~\ref{prop:ipw}]  
We verify below the results about $\tipw(\xxi)$ and $\tau_{\ipw\mid G =g} \ (g= 0,1)$, respectively.

\paragraph*{\underline{Results about $\tau_{\text{\tiny IPW}}(\xxi)$.}}
The law of iterated expectations ensures 
\beginy\label{eq:ipw_1}
\E\left[\frac{\gi }{e(\xxi)}\cdot \dyi \mid \xxi \right] 
&=& \E\left[\frac{\gi }{e(\xxi)}\cdot \dyi \mid \gio , \xxi \right] \cdot \pr(\gio \mid \xxi) \nonumber\\
&=& \E[ \dyi \mid \gio , \xxi ].
\endy
By symmetry, we have
\beginy\label{eq:ipw_0}
\E\left[\frac{1-\gi}{1-e(\xxi)}\cdot \dyi \mid \xxi \right] 
= \E[ \dyi \mid \giz, \xxi ].
\endy
Plugging \eqref{eq:ipw_1}--\eqref{eq:ipw_0} in the definition of $\tipw(\xxi)$ ensures 
\begina
\tipw(\xxi) 
&=& \E\left[\frac{\gi }{e(\xxi)}\cdot \dyi - \frac{1-\gi }{1-e(\xxi)} \cdot \dyi\mid \xxi \right]\\
&=& \E\left[\frac{\gi }{e(\xxi)} \cdot \dyi\mid \xxi \right] - \E\left[\frac{1-\gi}{1-e(\xxi)} \cdot \dyi\mid \xxi \right] \\
&\overset{\eqref{eq:ipw_1}+\eqref{eq:ipw_0}}{=}& \E[ \dyi \mid \gio , \xxi ] - \E[ \dyi \mid \giz, \xxi ]\\
&=& \tdxx . 
\enda

\paragraph*{\underline{Results about $\tau_{\text{\tiny IPW}\mid G = g}$.}}
We verify in the following 
\beginy\label{eq:ipw_goal}
\tipwo  = \E[\tdxx  \mid \gio ],
\endy where  
\beginy\label{eq:tipwo_def}
\tau_{\ipw \mid G_i = 1} = \E[\dyi \mid \gio ] - \E\left[\dfrac{e(\xxi)}{e} \cdot \dfrac{1-\gi}{1-e(\xxi)} \cdot \dyi \right]. 
\endy
That $\tipwz = \E[\tdxx  \mid \giz  ]$ follows by symmetry. 

First, the right-hand side of \eqref{eq:ipw_goal} equals 
\begina
\E[\tdxx  \mid \gio  ]
&=& \E\Big\{ \E[\dyi \mid \gio , \xxi ] - \E[\dyi \mid \giz, \xxi ] \mid \gio  \Big\}\\
&=& \E\Big\{ \E[\dyi \mid \gio , \xxi ]\mid \gio  \Big\} - \E\Big\{\E[\dyi \mid \giz, \xxi ] \mid \gio  \Big\}\\
&=&  \E[\dyi \mid \gio ] - \E\Big\{\E[\dyi \mid \giz, \xxi ] \mid \gio  \Big\}.
\enda 
This, together the definition of $\tipwo$ in \eqref{eq:tipwo_def}, ensures \eqref{eq:ipw_goal} equals
\beginy\label{eq:ipw_goal2}
\E\left[\dfrac{e(\xxi)}{e} \cdot \dfrac{1-\gi}{1-e(\xxi)} \cdot \dyi \right] = \E\Big\{\E[\dyi \mid \giz, \xxi ] \mid \gio  \Big\}. 
\endy 
We show \eqref{eq:ipw_goal2} in the following. First, the law of iterated expectations ensures
\beginy\label{eq:ipw_att_ss1}
\E\Big[ (1-\gi) \cdot \dyi \mid \xxi \Big] 
&=&
\E[ \dyi \mid \giz, \xxi ] \cdot \pr(\giz   \mid \xxi) \nonumber\\
&=& \E[ \dyi \mid \giz, \xxi ] \cdot \left\{1-e(\xxi)\right\}.
\endy 
This ensures the conditional version of the left-hand side of \eqref{eq:ipw_goal2} equals 
\beginy\label{eq:ipw_c}
\E\left[\dfrac{e(\xxi)}{e} \cdot \dfrac{1-\gi}{1-e(\xxi)} \cdot \dyi \mid \xxi \right]
&=& 
\dfrac{e(\xxi)}{e} \cdot \dfrac{1}{1-e(\xxi)} \cdot 
\E\Big[ (1-\gi) \cdot \dyi \mid \xxi \Big] \nonumber \\
&\overset{\eqref{eq:ipw_att_ss1}}{=}& \dfrac{e(\xxi)}{e} \cdot \dfrac{1}{1-e(\xxi)} \cdot \E[ \dyi \mid \giz, \xxi ] \cdot \left\{1-e(\xxi)\right\} \nonumber\\
&=& \dfrac{e(\xxi)}{e}  \cdot \E[ \dyi \mid \giz, \xxi ].   
\endy
In addition, $\E[\dyi\mid \giz, \xxi]$ is a function of $\xxi$ so that 
\beginy\label{eq:ipw_ss}
\E(\gi \mid \xxi)  \cdot \E[ \dyi \mid \giz, \xxi ] = 
\E\Big( \gi \cdot \E[ \dyi \mid \giz, \xxi ] \mid \xxi \Big). 
\endy 
This, together with \eqref{eq:ipw_c} and the law of iterated expectations, ensures \eqref{eq:ipw_goal2} as follows: 
\begina
\E\left[\dfrac{e(\xxi)}{e} \cdot \dfrac{1-\gi}{1-e(\xxi)} \cdot \dyi \right]
&=& 
\E\left\{ \E\left[\dfrac{e(\xxi)}{e} \cdot \dfrac{1-\gi}{1-e(\xxi)} \cdot \dyi \mid \xxi \right] \right\}\\
&\overset{\eqref{eq:ipw_c}}{=}& 
\E\left\{ \dfrac{e(\xxi)}{e}  \cdot \E[ \dyi \mid \giz, \xxi ]   \right\}\\
&=& 
e^{-1} \cdot \E\Big\{ \E(\gi \mid \xxi)  \cdot \E[ \dyi \mid \giz, \xxi ]   \Big\}\\
&\overset{\eqref{eq:ipw_ss}}{=}& 
e^{-1} \cdot \E\left\{ \E\Big(\gi \cdot \E[ \dyi \mid \giz, \xxi ]\mid \xxi\Big) \right\}\\
&=&  
e^{-1} \cdot \E\Big\{ \gi  \cdot \E[ \dyi \mid \giz, \xxi ]  \Big\}\\
&=& 
\E\Big\{ \E[ \dyi \mid \giz, \xxi ] \mid \gio  \Big\}.
\enda
\end{proof}

\clearpage

\section{Additional Information on the Application}\label{sec:sm.app}

Below we provide additional information on the empirical application. 
\begin{itemize}\itemsep0em
    \item Table~\ref{table:stats}: Descriptive statistics
    \item Figure~\ref{fig:ps}: Assessing Assumption~\ref{assm:overlap} (overlap) using the estimated propensity score
    \item Figure~\ref{fig:sens}: Sensitivity analysis for the relationship between $G$ and $\Delta Y$
\end{itemize}
\clearpage

\begin{table}[ht]
\renewcommand{\arraystretch}{0.7}
\centering
\caption{Descriptive Statistics}\label{table:stats}
\vspace{-0.5em}
\begin{tabular}{lrrrrrr}
  \hline\hline
Variable & N & Mean & Median & SD & Min & Max \\ 
  \hline
  High social capital & 921 & 0.50 & 1.00 & 0.50 & 0.00 & 1.00 \\ 
  Per capita grain production & 921 & 285.39 & 272.14 & 96.35 & 69.72 & 613.36 \\ 
  Ratio of non-farming land & 921 & 0.22 & 0.21 & 0.09 & 0.01 & 0.48 \\ 
  Share of urban population & 921 & 6.86 & 5.62 & 4.54 & 0.56 & 25.93 \\ 
  Distance from Beijing (km) & 921 & 937.29 & 808.83 & 563.43 & 43.11 & 2149.29 \\ 
  Distance from provincial capital (km) & 921 & 194.90 & 182.51 & 101.94 & 0.00 & 481.79 \\ 
  Suitable for rice cultivation & 921 & 0.42 & 0.00 & 0.49 & 0.00 & 1.00 \\ 
  Share of ethnic minorities & 921 & 0.17 & 0.00 & 0.37 & 0.00 & 1.00 \\ 
  Average years of education & 921 & 2.30 & 2.26 & 0.63 & 0.72 & 4.37 \\ 
  Log population & 921 & 12.49 & 12.57 & 0.72 & 9.59 & 13.95 \\ 
  Mortality rate (‰) in 1954 & 921 & 12.68 & 12.38 & 4.91 & 0.30 & 92.56 \\ 
  Mortality rate (‰) in 1955 & 921 & 12.47 & 11.88 & 4.35 & 1.36 & 46.94 \\ 
  Mortality rate (‰) in 1956 & 921 & 11.92 & 11.61 & 3.90 & 1.26 & 39.20 \\ 
  Mortality rate (‰) in 1957 & 921 & 11.69 & 11.41 & 2.94 & 1.36 & 22.64 \\ 
  Mortality rate (‰) in 1958 & 921 & 13.65 & 11.86 & 6.53 & 1.11 & 47.32 \\ 
  Mortality rate (‰) in 1959 & 921 & 19.65 & 14.38 & 16.53 & 1.15 & 183.47 \\ 
  Mortality rate (‰) in 1960 & 921 & 29.51 & 19.47 & 26.84 & 1.15 & 218.88 \\ 
  Mortality rate (‰) in 1961 & 921 & 17.49 & 14.00 & 11.30 & 1.16 & 86.49 \\ 
  Mortality rate (‰) in 1962 & 921 & 11.01 & 10.40 & 4.01 & 1.43 & 51.49 \\ 
  Mortality rate (‰) in 1963 & 921 & 11.48 & 10.99 & 3.73 & 1.24 & 63.00 \\ 
  Mortality rate (‰) in 1964 & 921 & 12.82 & 11.81 & 3.91 & 1.39 & 28.96 \\ 
  Mortality rate (‰) in 1965 & 921 & 10.50 & 9.99 & 2.91 & 1.30 & 21.45 \\ 
  Mortality rate (‰) in 1966 & 921 & 9.86 & 9.48 & 2.71 & 1.18 & 24.52 \\ 
   \hline
  \multicolumn{7}{p{0.95\textwidth}}{\footnotesize\textbf{Note}: High social capital is defined as a county having a number of genealogy books greater than or equal to the sample median.} 
\end{tabular}
\end{table}
\clearpage

\begin{figure}[!ht]
    \centering
    \includegraphics[width=0.7\linewidth]{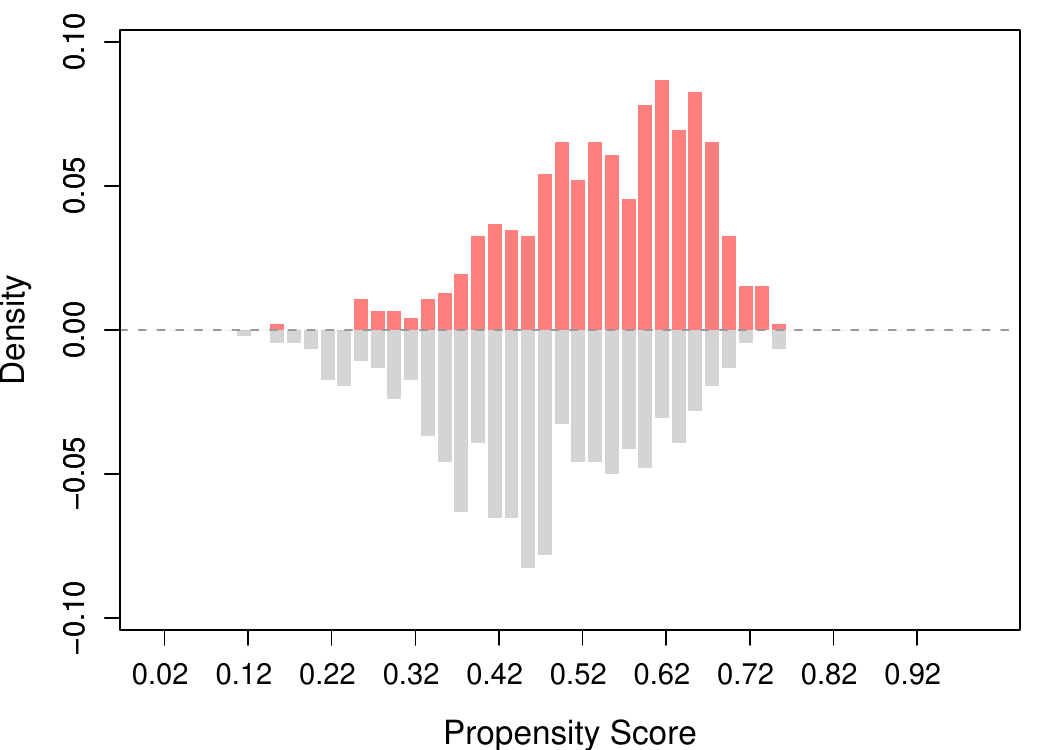}
    \caption{Accessing Assumption~\ref{assm:overlap} (overlap) using the estimated propensity score. Histograms of propensity scores estimated using a generalized random forest, based on the same set of covariates as in the regression analysis in Table~\ref{table:famine} columns (2)--(3), are shown in red for the group $\{i: G=1\}$ and gray for the group $\{i: G=0\}$.}
    \label{fig:ps}
\end{figure}

\clearpage

\begin{figure}[!ht]
    \centering
    \includegraphics[width=0.45\linewidth]{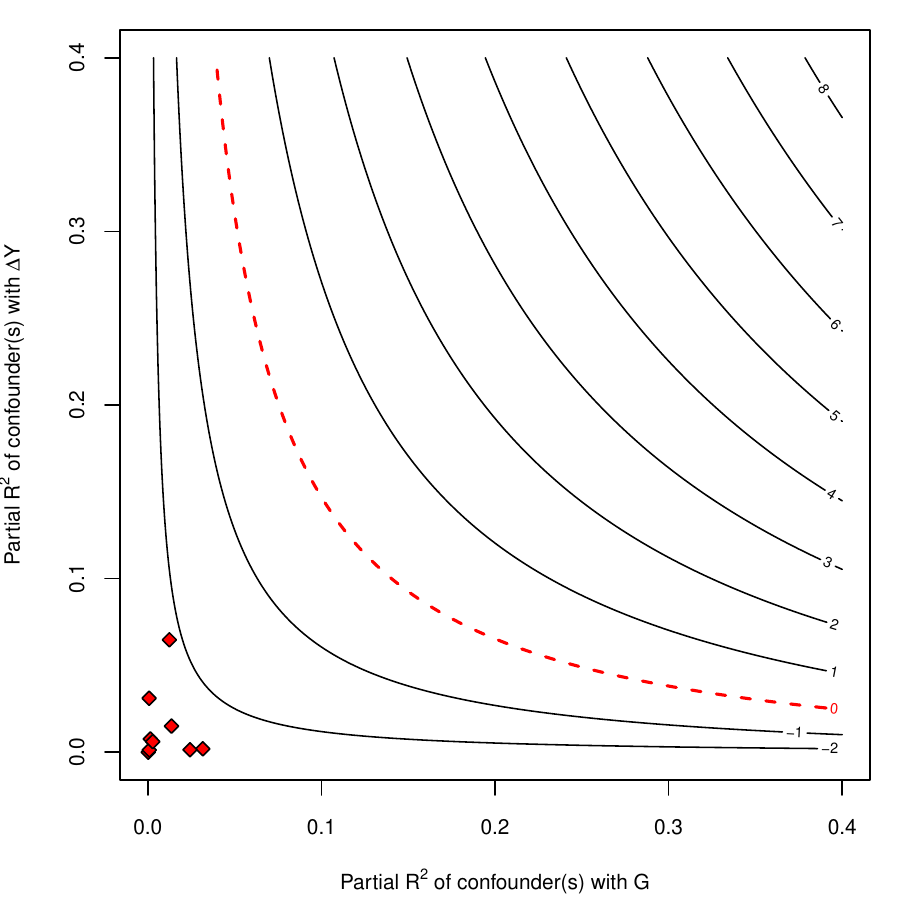}\hspace{1em}
    \includegraphics[width=0.45\linewidth]{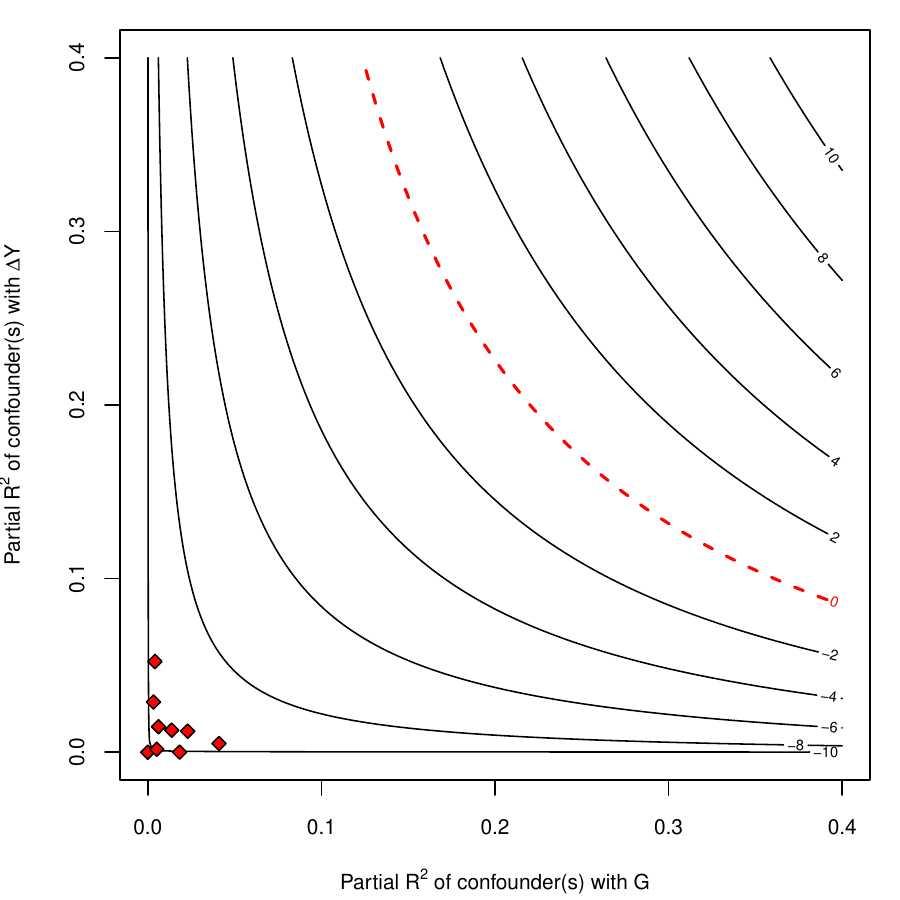}
    \caption{Sensitivity analysis based on \citet{cinelli2020making}. The left and right panels use binary and continuous measures of social capital $G$, respectively, with {\olsa} as the estimator. The outcome $\Delta Y$ is defined as the average mortality from 1958–1961 (the famine year) minus the 1957 mortality rate (baseline). The results show that a confounder whose associations with $G$ and $\Delta Y$ are comparable to those of existing pre-event covariates would not be strong enough to eliminate the negative association between $G$ and $\Delta Y$. The red diamonds in each plot illustrate the correlation of an existing covariate with both $G$ and $\Delta Y$.}
    \label{fig:sens}
\end{figure}
\clearpage

\fi} 

\end{document}